\documentclass[10pt]{article}

\usepackage{epsfig}
\usepackage{amssymb}
\usepackage{amsmath}

\renewcommand{\baselinestretch}{2.0}

\begin{document}

\pagestyle{plain}
\pagenumbering{arabic}
\setcounter{page}{1}


\begin{center}

{\Large \bf Progressive Processing of Continuous Range Queries in
Hierarchical Wireless Sensor Networks}

Jeong-Hoon Lee$^{*}$, Kyu-Young Whang$^{*}$, Hyo-Sang Lim$^{*}$, Byung-Suk Lee$^{**}$, and Jun-Seok Heo$^{*}$ \\
\vspace*{-0.20cm}
$^{*}$Department of Computer Science \\
\vspace*{-0.20cm}
Korea Advanced Institute of Science and Technology\,(KAIST) \\
\vspace*{-0.20cm}
$^{**}$Department of Computer Science \\
\vspace*{-0.20cm} University of Vermont, Burlington, VT 05405,
USA \\
\vspace*{-0.20cm} e-mail:\,$^{*}$\{jhlee, kywhang, hslim,
jsheo\}@mozart.kaist.ac.kr, $^{**}$bslee@cems.uvm.edu \\
\end{center}


\renewcommand{\baselinestretch}{1.6}
\begin{abstract}
{\small



In this paper, we study the problem of processing continuous range
queries in a hierarchical wireless sensor network. Recently, as the
size of sensor networks increases due to the growth of ubiquitous
computing environments and wireless networks, building wireless
sensor networks in a \emph{hierarchical} configuration is put forth
as a practical approach. Contrasted with the traditional approach of
building networks in a ``flat'' structure using sensor devices of
the same capability, the hierarchical approach deploys devices of
higher capability in a higher tier, i.e., a tier closer to the
server. While query processing in flat sensor networks has been
widely studied, the study on query processing in hierarchical sensor
networks has been inadequate. In wireless sensor networks, the main
costs that should be considered are the energy for sending data and
the storage for storing queries. There is a trade-off between these
two costs. Based on this, we first propose a \emph{progressive
processing} method that effectively processes a large number of
continuous range queries in hierarchical sensor networks. The
proposed method uses the query merging technique proposed by Xiang
et al. as the basis and additionally considers the trade-off between
the two costs. More specifically, it works toward reducing the
storage cost at lower-tier nodes by merging more queries, and toward
reducing the energy cost at higher-tier nodes by merging fewer
queries (thereby reducing ``false alarms''). We then present how to
build a hierarchical sensor network that is \emph{optimal} with
respect to the weighted sum of the two costs. It allows for a
cost-based systematic control of the trade-off based on the relative
importance between the storage and energy in a given network
environment and application. Experimental results show that the
proposed method achieves a near-optimal control between the storage
and energy and reduces the cost by $0.989~\sim~84.995$ times
compared with the cost achieved using the flat (i.e.,
non-hierarchical) setup as in the work by Xiang et al.

}
\end{abstract}
\renewcommand{\baselinestretch}{2.0}

\newtheorem{definition}{\bf Definition}
\newtheorem{strategy}{\bf Strategy}
\newtheorem{example}{\bf Example}
\newtheorem{property}{\bf Property}

\newenvironment{newidth}[2]{
 \begin{list}{}{
  \setlength{\topsep}{0pt}
  \setlength{\leftmargin}{#1}
  \setlength{\rightmargin}{#2}
  \setlength{\listparindent}{\parindent}
  \setlength{\itemindent}{\parindent}
  \setlength{\parsep}{\parskip}
 }
\item[]}{\end{list}}

%
%
\section{Introduction}\label{sec:Introduction}

As the computing environment evolves toward ubiquitous computing,
there has been increasing attention and research on sensor networks.
In the sensor networks environment, sensor nodes are connected
through the network to the server (or base station) which collects
data sensed at the nodes\cite{Abi05}. Example applications in such
an environment include environment monitoring(e.g., temperature,
humidity), manufacturing process tracking, traffic monitoring, and
intrusion detection in a surveillance system.

In particular, as \emph{wireless} network becomes more common, there
has been a lot of research on wireless sensor networks in which
sensor nodes are connected in an ad-hoc network configuration in
order to reduce the cost of deployment. In general, the objective in
a wireless sensor network is to deploy cheap sensor nodes with
limited resources (e.g., battery power, storage space) effectively
and to collect data from those sensor nodes by using their limited
resources efficiently \cite{Mad05}.

There is an increasing trend lately toward \emph{large-scale}
wireless sensor networks\cite{Pal07,Par06}, as the scope of
applications extends to municipality management, global
environmental monitoring, etc. These networks typically aim at
supporting a large number of sensor nodes deployed in a large area
for use by a large number of users. For example, in the Network for
Observation of Volcanic and Atmospheric Change (NOVAC)
project\cite{NOV07}, a wireless sensor networks deployed in 15
volcanoes spread across five continents are connected in a
multi-tier configuration to support a global volcano monitoring
project. As another example, the EarthNet Online\cite{Ear07}
collects earth observation information such as the worldwide weather
and bird migrations through wireless sensor networks and makes the
information available for thousands of individuals or organizations.
This kind of scale upgrade will bring about a proportionate increase
of the number of concurrent queries and the amount of sensor data.
Thus, we expect an increasing importance of processing a large
number of queries and a high volume data \emph{effectively} in
wireless sensor networks. In addition, we expect that building such
large scale wireless sensor networks \emph{economically} is
important as well.

With these regards, in this paper, we consider \emph{storage}
requirement needed to store queries in sensor nodes and
\emph{energy} consumption (i.e., battery capacity) needed to send
the collected data from those nodes to the server. There exists a
trade-off between these two cost factors. Let us explain this
trade-off with the centralized approach and the distributed
approach\cite{Rat03}, which are the two naive approaches to build
wireless sensor networks. In the centralized approach, the sensor
nodes do not store any query and simply send all data to the server,
which then processes all the queries on the data received. In this
case, there is no storage cost to store queries in individual sensor
nodes but the energy cost is very high. In the distributed approach,
on the other hand, individual sensor nodes store all the queries and
send only the results of processing the queries to the server, which
then simply collects the received query results (This scheme is
known as \emph{in-network query processing}\cite{Yao02}). In this
case, the energy cost can be reduced but the storage cost is high.

\vspace*{1pt}Neither of these two approaches is suitable for
building \emph{large scale} sensor networks. In the centralized
approach, since data are accumulated over the course of being
relayed toward the server, sensor nodes near the server should send
more data than the nodes farther from the server. As the number of
nodes increases, this phenomenon will become more serious. In other
words, sensor nodes closer to the server consume more energy than
other nodes farther from the server -- for sending not only the data
generated by themselves but also the data received from other nodes;
as a result, those nodes will be burnt out within a short time.
Thus, the centralized approach is not appropriate for large scale
sensor networks. On the other hand, the distributed approach becomes
infeasible as the number of queries increases. A sensor node is not
able to process a large number of queries due to the limitation on
its memory and computing power. Consider as an example inexpensive
Micamotes\cite{Par06}, which typically have only 8$\sim$128~Kbyte
flash memory and 0.5$\sim$8~Kbyte RAM. Suppose a mote has 64Kbyte
flash memory and 10\% of it is available for storing two-dimensional
range queries. Additionally, suppose that each attribute value of a
query is a real number of four bytes long and that the selection
condition of a query is expressed as \emph{$c_1$} \emph{$op_1$}
\emph{A} \emph{$op_2$} \emph{$c_2$} (\emph{A}: attribute name;
\emph{$c_1$} and \emph{$c_2$}: attribute values; \emph{$op_1$} and
\emph{$op_2$}: binary comparison operators). Then, the size of one
query is at least 16 bytes\cite{Mad05}.
 and, thus at most 400 queries can be
stored in one mote. Obviously, these motes are far too short to
store thousands of queries expected of large scale networks.
Upgrading the sensor nodes to those with large enough memory will
raise the expense, which is not acceptable when there are so many
sensor nodes to be deployed.

\vspace*{1pt}Recently, in order to overcome these large scale
problems, building wireless sensor networks in a \emph{hierarchical}
configuration is considered a practical alternative. A hierarchical
wireless sensor network is organized in a \emph{multi-tier
architecture}\cite{Aky07} configured with sensor nodes having
different amounts of resources and computation power. Nodes closer
to the server have more resources and computation power than those
farther from the server, and this makes it possible to carry out the
processing that cannot be done with low-capacity nodes only. In
hierarchical wireless sensor networks, nodes with smaller resources
and computing power are recursively connected to nodes with more
resources and computing power\cite{Qin06,Sin03,Aky07}; thus, nodes
near the server are capable of handling the larger amount of data
accumulated from lower tiers. We think this configuration is
suitable for resolving the query processing problem in large-scale
networks mentioned above. Currently, however, the main stream of
research on wireless sensor network query processing is for
\emph{flat} sensor networks (i.e., sensor networks that consist of
nodes with the same capability). Accordingly, research on query
processing for hierarchical sensor networks has been less than
adequate.

\vspace*{1pt}This paper proposes a method for building large scale
hierarchical sensor networks to process queries effectively with
respect to the trade-off between the energy cost and the storage
cost. The queries considered in this paper are \emph{continuous
range} queries. Range queries are an important query type in many
sensor network applications, particularly in monitoring
applications\cite{Mad05}, and there has been active research done to
improve range query processing performance\cite{Li03}. The method
proposed in this paper is based on the technique of systematically
controlling the trade-off between the energy cost and the storage
cost through controlled merging of queries with similar ranges.
There are existing methods proposed to reduce the energy cost by
merging queries to avoid duplicate transmission of query
results\cite{Mul06,Xia06,Xia07}. They, however, all focus on
\emph{flat} sensor networks and, therefore, cannot utilize the
characteristics of hierarchical sensor networks in which nodes at
different tiers have different capabilities. Besides, their work
does not reflect anything about the trade-off because they do not
consider the storage cost at all. In contrast, in this paper, we
fully utilize the characteristics by employing a \emph{progressive}
approach, which merges increasingly more queries as the tier goes
from the server toward the lowest tier and, in this way, finds the
optimal merging at each tier in consideration for the trade-off.
More specifically, at lower tier nodes, which are larger in number,
the approach works toward reducing the storage requirement by
reducing the number of queries through more aggressive merging; in
contrast, at higher tier nodes, which are smaller in number, the
approach works toward storing more queries through less aggressive
merging and, in return, reducing the energy consumption by
increasing the query accuracy by filtering out more unnecessary
data.

\vspace*{1pt}In this paper, we first propose the model and
algorithms of the progressive query processing method. This method
has two phases: \emph{query merging} and \emph{query processing}.
The key idea in the query merging phase is to merge queries
progressively as the tier goes from the highest (i.e., the server)
to the lowest. In other words, it merges the input queries to
recursively generate queries to be stored at the next tier nodes,
first merging the input queries to generate queries for the second
tier nodes, and then merging them to generate the ones for the third
tier nodes, and so on. We say that the queries thus stored at
multiple tiers form the \emph{inverted hierarchical query
structure}\footnote{It is a \emph{forest} structure to be more
precise (see Figure\,\ref{fig:e_merging_sample}).} as a whole.

The \emph{Inverted hierarchical query structure} is a new structure
proposed in this paper. It is built from a multi-dimensional index
storing the query ranges, by partitioning the index into multiple
levels and then storing the root level of the index at the
lowest-tier sensor nodes and the leaf level of the index in the
server. This structure is based on the characteristics of
hierarchical sensor networks that sensor nodes at a higher level
store more detailed information while sensor nodes at a lower level
store more abstract information. Thus, the structure is an inverse
of a general tree-like index structure.

In the query processing phase, the queries are processed
progressively, that is, by refining the query result to be more
accurate as data are sent from a lower tier to a higher tier. For
this, the inverted hierarchical query structure is used to retrieve
the query result at each tier.

Next, we propose a method that builds an optimal hierarchical sensor
network by systematically controlling the trade-off between the
storage cost and the energy cost according to their weights. Since
the relative importance between the two costs may vary depending on
the application and environment, we formulate the cost of building
the network as a weighted sum of the two costs and minimize the
total cost. As the optimization target parameter, we use the
\emph{optimal merge rate} -- the average rate of merging queries at
each tier.

Finally, we show through experiments that the proposed method is
useful for building a hierarchical sensor network in a cost
effective manner. Specifically, first we show that there is little
difference between the optimal merge rate obtained from an analytic
model and the rate obtained from experiments; second, we show the
superiority of the proposed method over the existing query
processing method for flat sensor networks in terms of the total
cost.

The rest of this paper is organized as follows. Section 2 discusses
related work. Section 3 describes the model and the algorithms of
the proposed progressive processing method for hierarchical sensor
networks. Section 4 proposes an analytical method for effectively
building a hierarchical sensor network. Section 5 shows the
superiority of the proposed method over the existing method through
experiments. Section 6 concludes the paper.
\vspace*{-0.30cm}
\section{Related Work}
\vspace*{-0.30cm}

\noindent In this section, we review the existing research on the
continuous range query processing in sensor networks and the state
of the art in the \emph{hierarchical} wireless sensor networks.

\vspace*{-7pt}\subsection{Continuous range query processing in
sensor networks}\vspace*{-3pt}

\noindent In sensor networks, range query processing can be
classified into single range query processing and multiple range
query processing. Single range query processing executes only one
range query in a system. Multiple range query processing
concurrently executes many range queries in a system.

\vspace*{5pt} \noindent \textbf{Single continuous range query
processing}\vspace*{4pt}

\noindent Li et al.~\cite{Li03} apply the data-centric storage to
continuous single query processing. The query processing using the
data-centric storage runs as follows. For storing data, each sensor
node sends collected data to sensor nodes, where the target sensor
nodes are determined by the value of the data element. For
processing queries, the server sends a query to only those sensor
nodes that have the result data of the query. In the same work, Li
et al. study an index structure using an order-preserving hash
function for distributing data. That is, nodes that are physically
adjacent have the adjacent value ranges of data stored in the
nodes.As a result, the method reduces the query processing cost by
reducing the average number of hops for sending queries and query
results. Madden et al.\cite{Mad05} consider storing data in local
sensors (unlike the data-centric approach) and propose building an
R-tree-like index (called \emph{SRTree}) based on the range of
sensing values. Both of these works focus on single query
processing. Hence, they are not applicable for recent query
processing environments that register many queries and process them
concurrently.

\vspace*{5pt}\noindent \textbf{Multiple continuous range query
processing}\vspace*{4pt}

\noindent Ratnasamy et al.~\cite{Rat03} propose two basic query
processing approaches for multiple query processing in wireless
sensor networks. One approach processes queries at the server(called
the \emph{centralized} approach), and the other approach processes
queries at the sensor node(called the \emph{distributed} approach).
In the former approach, all queries are stored in the server, and
the sensor nodes send all sensed data to the server for query
processing. This approach is effective only if the size of the
region equivalent to the union of all query regions is close to the
size of the entire domain space and, otherwise, incurs the overhead
of sending unnecessary data to the server. This approach can reduce
the memory requirement of the sensor nodes because it does not store
any query in them, but has the disadvantage of incurring significant
energy consumption because all data must be sent to the server. In
the latter approach, each sensor node stores all queries
disseminated from the server and sends to the server only the result
of processing the sensor data. Thus, this approach may not have the
problem of the former approach, but has the disadvantage that the
sensor nodes may not be able to store all queries due to
insufficient memory if the number of queries is large. From these
two basic query processing approaches, we can observe that there is
a trade-off between the memory and the energy which are two
important resources of sensor nodes.

Furthermore, recently, there has been research to complement the
centralized approach and the distributed approach. Specifically, the
proposed methods are to share query processing in an overlapping
region in case there are overlapping query conditions. By
identifying the overlapping regions among the user queries and
rewriting the queries accordingly, the proposed methods eliminate
duplicate processing and duplicate data transmission. These methods
can be classified into the \emph{partitioning} method and the
\emph{merging} method.

In the partitioning method, the server partitions the individual
query regions into overlapping regions and non-overlapping regions.
Then, it sends the partitioned regions and the original queries to
sensor nodes, which store them. Query processing is done for each
partitioned region, and the query results are merged in the server
or sensor nodes. Trigoni et al.~\cite{Tri05} and Yu et
al.~\cite{Yu04} use this method to process range queries on the
\emph{location} information of sensor nodes. This method has the
advantage that the result of merging the results of processing each
partition is the same as the result of processing the original
queries and, therefore, no ``false alarm'' will happen. It, however,
has the disadvantage that, if there are a large number of
overlapping query conditions, then the number of partitions to be
stored in certain sensor nodes increases and, thus, the necessary
storage increases as well.

In the merging method, the server merges the regions of overlapping
queries into one merged query region. The server then sends the
merged queries to the sensor nodes that store them. Query processing
results are then ``reorganized'' into those of the original queries
in the server or sensor node. This method has the advantage that it
can process a large number of queries at the same time by reducing
the number of queries stored in a sensor node. It, however, has the
disadvantage that a ``false alarm'' may happen as a result of
merging queries. Muller and Alonso\cite{Mul06} propose a method that
compares the predicates of the range queries to extract those common
to all queries and generates one query that has only the common
predicates as the query condition. In this method, if there is no
predicate common to all queries, then one query with no query
condition is generated and, thus, has the problem of incurring a lot
of false alarms in that case. Xiang et al.~\cite{Xia06,Xia07}
propose a method which \emph{incrementally} merges overlapping query
regions and processes the resulting merged queries instead of the
original queries. Here, the incremental merging is done until the
cost of sending the false alarms occurring when queries are merged
is no larger than the cost of sending duplicate results of
overlapping query regions when queries are not merged. Xiang et
al.'s query processing method has the meaning of a hybrid approach
(i.e., reducing the needed memory amount and the data transmission
amount) taking advantage of both the centralized approach and the
distributed approach, but targets ``flat'' sensor networks in which
all sensor nodes in the network have the same capability and store
the same set of merged queries. Thus, this method has the problem
that it cannot utilize the characteristics of \emph{hierarchical}
sensor networks. Our method in this paper basically uses the same
query merging method as Xiang et al.'s, but enhances it to control
the rate of merging queries depending on the capabilities of
individual nodes and to build a hierarchical sensor network. Our
method has the advantage that it allows for a systematic control of
the trade-off between the memory amount needed and the amount of
data sent.

\vspace*{-0.1cm}\subsection{Hierarchical wireless sensor
networks}\vspace*{-0.1cm}

As the scale of sensor networks increases, the hierarchical
structure is used more in real applications than the flat structure
in which all sensor nodes have the same capability\cite{Aky07}.

Representative examples of such hierarchical wireless sensor
networks are PASTA(Power Aware Sensing, Tracking and
Analysis)\cite{Sin03} mentioned in COSMOS\cite{Sin03} and
SOHAN\cite{Gan04}. PASTA is used in military applications for enemy
movement surveillance and is configured with the server and about
400 intermediate tier nodes each clustering about 20 sensor nodes.
SOHAN is used in traffic congestion monitoring applications to
measure the traffic volume using roadside sensor nodes and is
configured with the server and about 50 intermediate tier nodes each
clustering about 200 sensor nodes.

We expect that hierarchical sensor networks will be increasingly
more utilized in the future as the scale and the requirement of
applications increase. However, there has not been any research done
on processing multiple queries talking advantage of the
characteristics that sensor nodes at different tiers have different
capabilities. Srivastava et al.~\cite{Sri05} investigated how and on
which node to process each operation during query processing in a
hierarchical sensor network. This research, however, mainly deals
with single query processing and, thus, is difficult to apply to
multiple query processing. In this paper, we propose a method for
processing multiple queries effectively by utilizing the
characteristics of hierarchical sensor networks, i.e., the
multi-tier structure made of sensor nodes with different resources
and computing power.

\vspace*{-0.1cm}
\section{Progressive processing in hierarchical wireless sensor networks}\label{sec:ProgressiveProcessing}
\vspace*{-0.1cm}

In this section, we present the progressive processing model and
algorithms in hierarchical (i.e., multi-tier) wireless sensor
networks.

\vspace*{-0.5cm}
\subsection{Overview}
\vspace*{-0.5cm}

In progressive processing, we systematically control the total
processing cost by having the larger number of lower-capacity nodes
(at lower tiers) partially process queries and the smaller number of
higher-capacity nodes (at higher tiers) process the remainder.

\begin{figure}[h!]
  \centering
  \includegraphics[width=15cm]{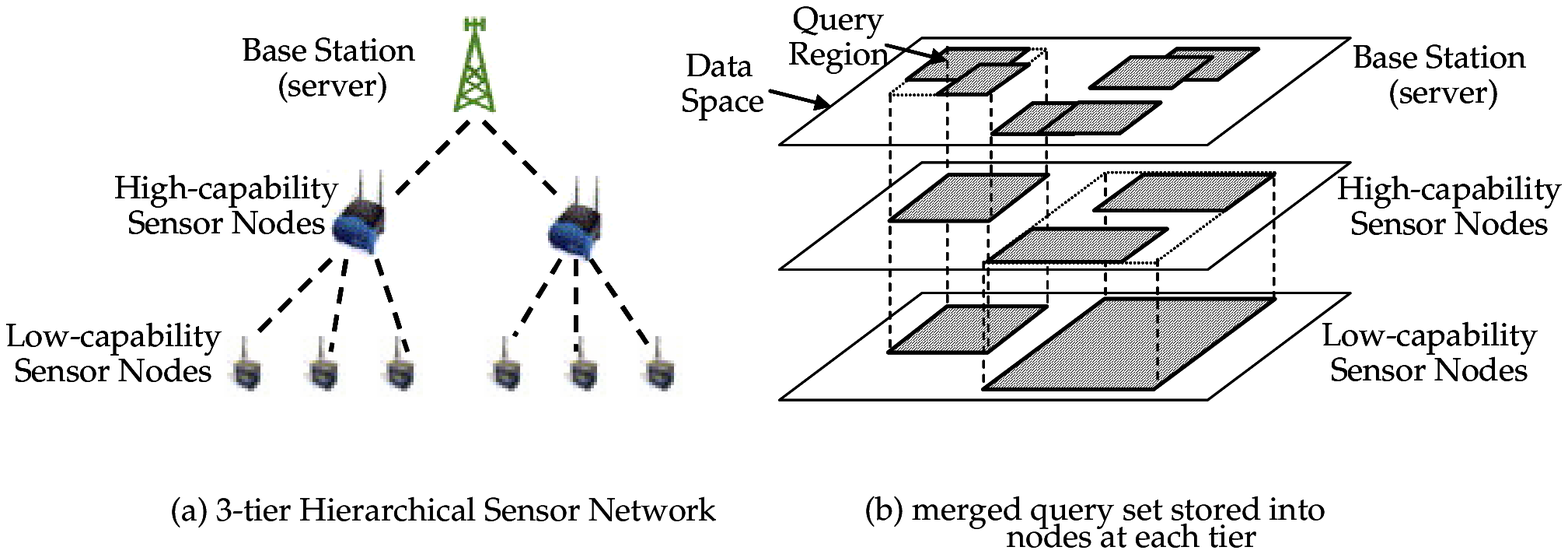}
  \centerline{\small This figure shows a three-tier network as an example.}
  \caption{Inverted hierarchical query structure in a hierarchical wireless sensor network.}
  \label{fig:e_inverted_tree}
\end{figure}

\vspace{0.3cm} \noindent \textbf{Example\,1}\,(Progressive
processing in hierarchical wireless sensor networks): {\rm
Figure\,~\ref{fig:e_inverted_tree}(a) shows an example of a
hierarchical sensor network organized in three tiers. The nodes at
the third (i.e., lowest) tier are the largest in number but the
smallest in capability and are connected to the more capable nodes
at the second tier. All nodes except the server generate data (i.e.,
partial query results) periodically and send them to the server
relayed via the nodes at higher tiers. The server then provides the
final query result to the user.
Figure\,~\ref{fig:e_inverted_tree}(b) shows the set of queries
stored in the nodes at each tier at the end of the query merging
phase. In this figure, the rectangular regions represent range
queries, and the boundary rectangle represents the domain space
defined by the attributes specified in the queries. The server
stores six original queries, the second tier nodes store three
queries resulting from the merge of the six original queries, and
the third tier stores two queries resulting from further merging
them. In the query processing phase, sensor nodes at the lowest tier
process the two queries on the sensed data and send to the second
tier only the data satisfying the conditions (i.e., ranges) of the
two queries. Then, the sensor nodes at the second tier process the
three queries on the data sent from nodes at the lower tier and the
data they generate on their own, and send to the server only the
data satisfying the query conditions. Since nodes at a higher tier
have queries of finer granularity, they can reduce ''false alarms''
and thereby reduce energy consumption. The server processes the
original queries on the data sent from all nodes at lower tiers and
provides the final result to the user.} \hfill\fbox{}

\vspace{0.3cm}From Figure\,\ref{fig:e_inverted_tree}(b), we can see
that the stored queries altogether form an \emph{inverted} structure
of a multi-dimensional index tree. In contrast to a
multi-dimensional index tree structure in which all objects are
stored in the leaf nodes and are merged to become more abstract at a
higher level, in the proposed structure, the root (i.e., server)
stores all objects (i.e., queries) and they are merged to become
more abstract at a lower level.

The progressive processing has the \emph{query merging} phase which
generates queries to be stored at each tier of the hierarchical
sensor network to form an inverted hierarchical query structure and
the \emph{query processing} phase which processes sensed data and
sends the result to the server using the inverted hierarchical query
structure. Query merging is performed off-line in batch processing,
and query processing is performed on-line every time data are
generated. In query merging, queries are sent toward the lowest tier
while merged ``progressively'', and, in query processing, the sensor
data are sent toward the server while being filtered
``progressively''.

In the query merging phase, minimum bounding rectangles (MBRs) are
obtained from the queries and expressed as merged queries. In this
case, it is important to decide how many MBRs the queries should be
merged into because the number of MBRs affects the trade-off between
the energy consumption and the storage usage. That is, if more
queries are merged, then the storage space used by the sensor nodes
to store queries is reduced, but the energy consumption is increased
due to more frequent false alarms. In this section, we present the
model and algorithms under the assumption that the number of merged
queries is known at each tier. Then, in section 4, we present a
method for determining the optimal number of merged queries
analytically using a cost model.

In the query processing phase, all sensor nodes except the server
process their own sensed data and the data received from the nodes
at lower tiers, and send the results to the nodes at the next higher
tier. Since more queries (of finer granularity) are stored at the
higher tier nodes, the accuracy of query result is higher in them,
thus generating the query result progressively.

\subsection{Network and data models}
In this section, we first define the hierarchical sensor network.
Then, we explain data and queries used in this paper.

\vspace*{5pt}\noindent \textbf{The hierarchical sensor
network}\vspace*{4pt}

\noindent We make the following assumption about the configuration
of a hierarchical sensor network. All sensor nodes are connected to
form a tree rooted at the server, and the nodes at the same depth
make one tier. Data are generated by not only the nodes at the
lowest tier but also those at intermediate tiers, and the sensed
data are sent to the server though the nodes at higher tiers. All
sensor nodes at the same tier have the same capability, that is, the
same amount of memory and battery power. Nodes closer to the server
have higher capability, that is, a larger amount of memory and
battery power. In addition, all nodes at the same tier store the
same set of queries.

There have been various research on the hierarchical sensor network
in the literature. However, the definitions of the hierarchical
sensor network vary depending on specific environments.
Nevertheless, it is a common understanding that a hierarchical
sensor network consists of multiple tiers and deploys devices of
different capabilities at different tiers\cite{Gan04,Sri05,Aky07}.
We define the hierarchical sensor network as in Definition\,1.
\newpage
\begin{definition}[The hierarchical sensor
network]\label{def:hierarchical_sensor_network}{\rm The hierarchical
sensor network is defined as a tree $T=(V,E)$ of height $h$, where
$V$ is a set of vertices representing the sensor nodes and the
server in the network (the root represents the server), and $E$ is a
set of edges representing the direct connection between a sensor
node and its parent node. Let $node_i$ denote the node at $i^{th}$
tier (1 $\leq$ $i$ $\leq$ h). Let $s_i$ and $e_i$ denote the amount
of storage and the amount of energy of $node_i$, respectively. Then,
a hierarchical sensor network satisfies relationship: $s_j$ $>$
$s_k$ and $e_j$ $>$ $e_k$ (1 $\leq$ $j$ $<$ $k$ $\leq$ $h$). }
\hfill\fbox{}
\end{definition}

\vspace*{5pt}\noindent \textbf{Query and data}\vspace*{4pt}

\noindent In this paper, we focus on the range query as the query
type in the hierarchical sensor network since it is an important
query type in sensor networks
applications\cite{Li03,Mad05,Mul06,Xia06}. Consider a
multi-dimensional domain space defined by the query attributes.
Then, in the domain space, a query and a data element are
represented as a hyper\_rectangular region and a point,
respectively\cite{Lim06}.

\vspace*{-0.5cm}
\subsection{Progressive query merging}
\vspace*{-0.1cm}

\subsubsection{The model}

\begin{figure}[h!]
  \centering
  \includegraphics[width=5in]{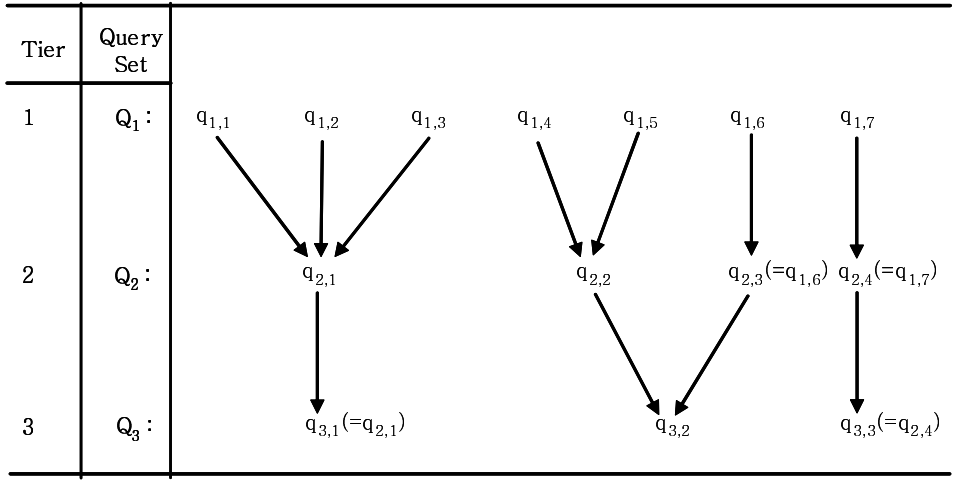}\\
  \caption{An example of progressive query merging.}\label{fig:e_merging_sample}
\end{figure}

\noindent Query merging in the first phase of progressive processing
is done by finding the MBR enclosing the queries to be merged.
\emph{Progressive} query merging means that more queries are merged
as the merging progresses to lower tiers. Thus, the size of a query
region is larger at a lower tier while the number of queries is
smaller. Let us refer to a query represented by an MBR that encloses
certain queries at a higher tier node as a \emph{merged query}, and
denote the set of queries (or the \emph{query set}) stored at the
$i^{th}$-tier node as $Q_i$. Then, we can represent the set of
merged queries at each tier as one level in the inverted
hierarchical query structure, as shown in
Figure\,\ref{fig:e_merging_sample}. In this figure, an arrow
represents the direction of query merging; queries at the tail of an
arrow are merged to the query at the head of the arrow. For
instance, the queries $q_{1,1}$,$q_{1,2}$ and $q_{1,3}$ at the
$1^{st}$ tier are merged to the query $q_{2,1}$ at the $2^{nd}$
tier.

\begin{figure}[h!]
  \centering
  \includegraphics[width=4.5in]{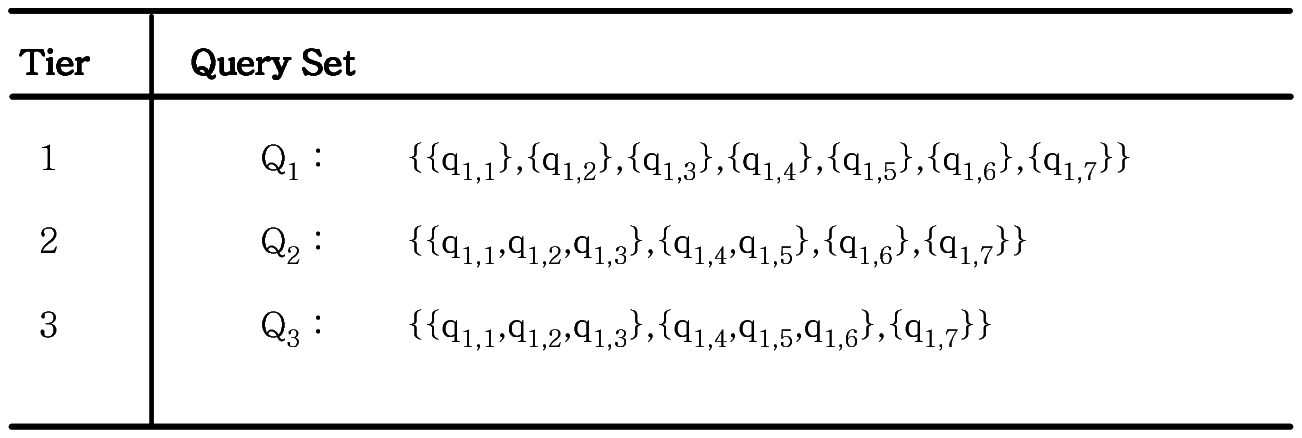}\\
  \caption{An example of progressive partition merging.}\label{fig:e_partition}
\end{figure}

The query merging can also be seen as merging the partition of a
disjoint set of queries. Figure\,\ref{fig:e_partition} illustrates
it with the same six queries as in
Figure\,\ref{fig:e_merging_sample}. The query $q_{2,1}$ in
Figure\,\ref{fig:e_merging_sample}, for example, corresponds to the
subset $\{ q_{1,1}, q_{1,2}, q_{1,3} \}$ of $Q_2$ in
Figure\,\ref{fig:e_partition}. The partitioning is coarser at a
lower tier.

\subsubsection{The algorithm}

\noindent For each $i^{th}$  tier, the progressive query merging
algorithm generates a merged query set $Q_i$ of a given size $C_i$.
The objective of the algorithm is to minimize the query processing
cost in consideration for the limited memory of sensor nodes. It is
difficult to predict the cost of query processing for a given set of
merged queries. The reason for this is that the cost depends not
only on the network-specific factors like routing but also on
unknown factors such as the query and data distributions. In this
paper, we use the simplified model proposed by Xiang et
al.\cite{Xia06}, in which the cost metric is the amount of data sent
during the query processing, as the basis and extend it to fit into
the hierarchical sensor network and take the memory usage into
consideration. In Xiang et al.'s model, the size $O$ of the
overlapping region among queries and the size $D$ of the dead region
(i.e., the region added in extra to make the MBR enclosing the merge
queries; it causes the false alarms) are calculated for each pair of
two queries that are candidates to be merged, and the pair that
maximizes the difference between the sizes of the two regions,
$O-D$, are merged. The effect of this is to merge queries with large
overlapped regions, which is a reasonable strategy for reducing the
data transmission cost.

The proposed algorithm performs the query merging using a greedy
approach based on the same strategy. Let $O(q_i,q_j)$ be the size of
the overlapping region between two queries $q_i$ and $q_j$, and
$D(q_i,q_j)$ be the size of dead region between them. The algorithm
chooses two queries $q_i$ and $q_j$ with the largest $O(q_i,q_j) -
D(q_i,q_j)$ from the set of queries that are either merged queries
or the original queries and merge them first. This strategy is the
same as the strategy used by Xiang et al.\cite{Xia06} except that
they consider only the pairs that satisfy $O(q_i,q_j) - D(q_i,q_j)
\geq 0$. Specifically, in consideration of the storage cost for
storing queries and the energy cost for sending query results, our
approach determines the fixed number of queries that are to be
stored into a sensor node at each tier. Then, we merge queries using
a greedy method until we reach the number while Xinag et al.'
approach determines the number of queries to be stored so as to only
minimize the amount of data sent.

\begin{figure}[h!]
  \centering
  \includegraphics[width=5.5in]{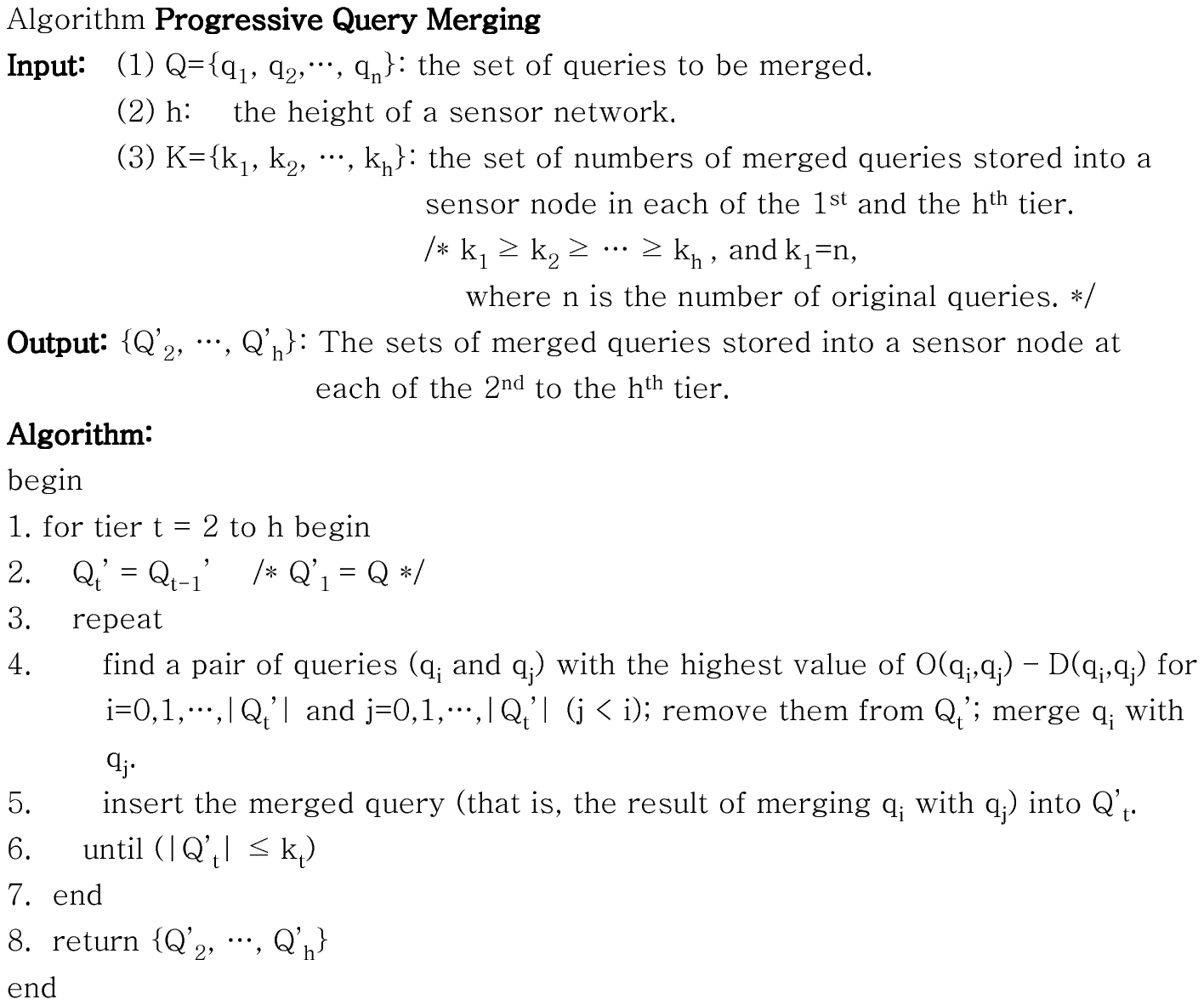}\\
  \caption{The progressive query merging algorithm.}\label{fig:e_query_merging_alg}
\end{figure}

Figure\,\ref{fig:e_query_merging_alg} shows the progressive query
merging algorithm. Inputs to this algorithm are the set of the
original queries $Q$, the height $h$ of the hierarchical sensor
network to be built, and the set of the numbers of merged queries
$K$ to be stored in every node at each tier. The output is the sets
of merged queries that are stored in every node at each tier. At
each tier $t$, the algorithm repeats merging two queries at a time
until the number of merged queries falls lower than $k_t$ (lines
3-6). In order to find the pair of queries to be merged, it
calculates the difference between the overlapping region and the
dead region over every pair of queries and merges the pair with the
maximum difference (lines 4-5).

\subsection{Progressive query processing}

\subsubsection{The model}

\begin{figure}[h!]
  \centering
  \includegraphics[width=5in]{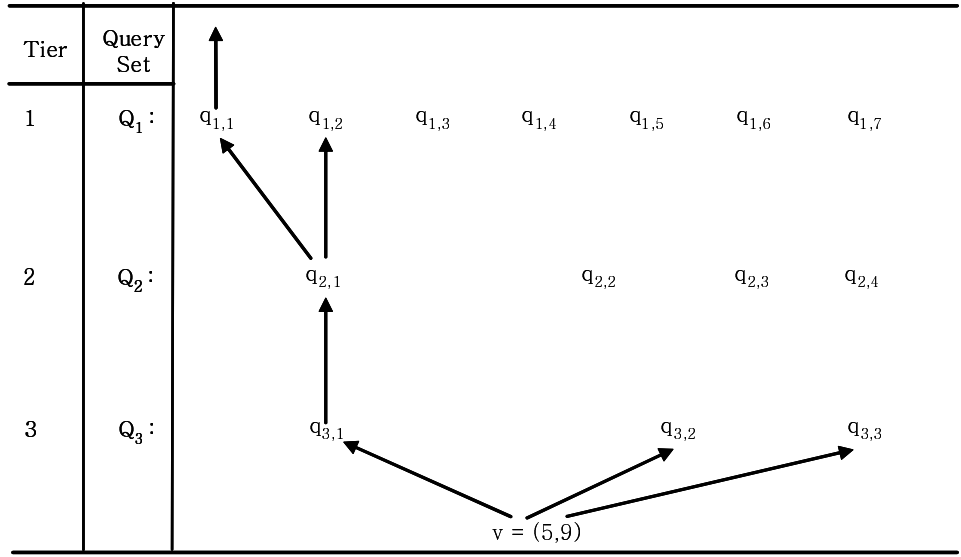}\\
  \caption{An example of progressive query processing.}\label{fig:e_processing_sample}
\end{figure}

\noindent In the query processing phase, for a given query, it is
decided whether a data element falls inside the query region, that
is, whether the attribute values representing the data element
satisfy the range predicates representing the region.
\emph{Progressive} query processing is the process of propagating
data elements bottom up in the inverted hierarchical query structure
from the lowest tier nodes to the highest tier node (server), while
filtering the data elements depending on the result of evaluating
the range predicates of the queries at each tier. (Precisely
speaking, multiple data elements are sent in a batch for the sake of
efficiency.)
Figure\,\ref{fig:e_processing_sample} shows an example of query
processing. In this figure an arrow denotes an upward flow of a data
item ($v$) as it satisfies the range predicate of the query at the
arrow tail. In this example, the query $q_{1,1}$ at the server
retrieves the data element $v$.

\subsubsection{The algorithm}

\begin{figure}[h!]
  \centering
  \includegraphics[width=6.5in]{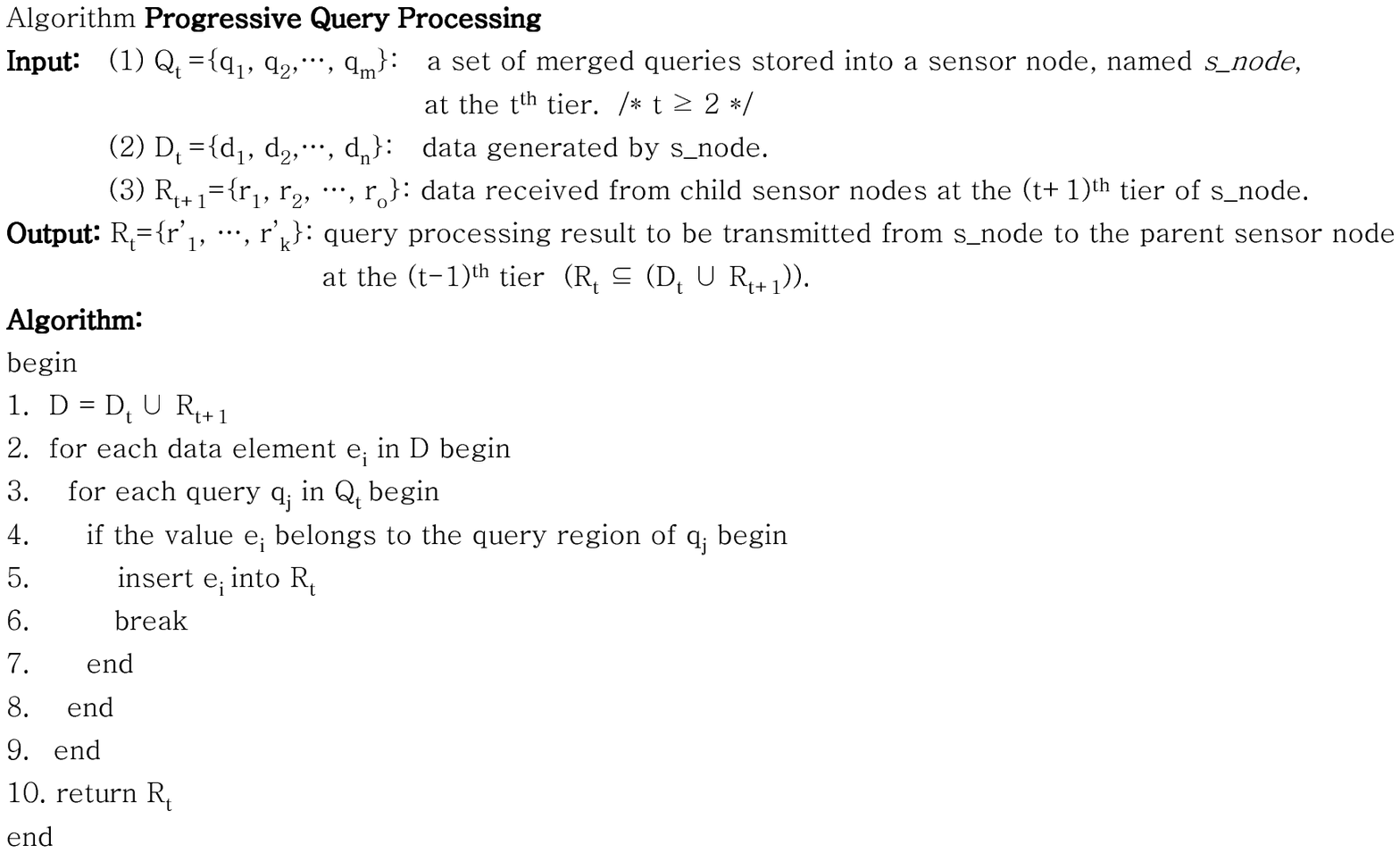}\\
  \caption{The progressive query processing algorithm.}\label{fig:e_query_processing_alg}
\end{figure}

\noindent Figure\,\ref{fig:e_query_processing_alg} shows the
progressive query processing algorithm. The algorithm is run
separatively at each tier of the hierarchical sensor network. The
algorithm is designed to run for each query on each data element,
which may not be the most efficient in terms of the query processing
time. However, the query processing time is independent of the
energy cost and the storage cost which are the main cost items
considered. Thus, it is not the focus of this paper.

In the progressive query processing, a sensor node at the $t^{th}$
tier($t\geq 2$) considers the data $D_t$ generated by itself and the
data $R_{t+1}$ resulting from the query processing at the
$(t+1)^{th}$ tier as the target data for query processing(line 1).
The node compares the set of merged queries $Q_t$ with the target
data and inserts only the data elements that satisfy the query
condition into $R_t$(lines 2-9). In order to prevent the node from
sending duplicate results of overlapping query regions among merged
queries, the algorithm stops the comparison once it finds a query
whose region contains the target data element(line 6)\footnote{When
the algorithm is run at the server, Line 6 should be removed because
the server must answer each query.}. Then, the node sends $R_t$ to
its parent node at the $(t-1)^{th}$ tier. This algorithm is run
separately in every node at each tier to progressively filter the
data to arrive at the highest tier (i.e., server). Finally, the
server(i.e., the $1^{st}$ tier) performs post-processing to select
the query results satisfying the condition of each query.

In this section, we have proposed the algorithms under the
assumption that the sensor nodes at each tier already know the
number of the merged queries to be stored. In the next section, we
propose an optimization method for determining the optimal number of
merged queries.

%
%
\section{Determining the Optimal Number of Merged Queries}\label{sec:DeterminingOptimal}
\vspace*{-0.30cm}

In this section, we propose an analytic method for determining the
optimal number of merged queries to be stored at each tier when
designing the hierarchical sensor network. We first propose the cost
model in Section~4.1 and then the cost optimization method in
Section~4.2.

\subsection{The cost model}

In this paper, we use the weighted sum of the storage cost for
storing queries and the energy cost for sending the query result as
the total cost. We use the total amount of memory used in all nodes
as the storage cost and the total amount of data sent during the
query processing as the energy cost. We use byte as the unit of both
the storage cost and the energy cost.

Eq.(\ref{eq:weightedSum}) shows the cost model expressed as the
function $weightd\_sum$.
\begin{eqnarray}
\mbox{\it Weighted\_Sum}& = & \alpha \cdot \mbox{\it the total
amount of data sent} +
\mbox{\it the total amount of memory used,}\nonumber \\%
& & \mbox{where $\alpha(>0)$ is the scale factor provided by the
user} \label{eq:weightedSum}
\end{eqnarray}
\vspace*{-1.0cm}

\noindent In this equation, the value of $\alpha$ indicates the
relative importance of the energy cost over the storage cost, and is
set by the user based on one's preference. That is, in the
environments where the energy cost is more important than the
storage cost, the user gives a larger value of $\alpha$, whereas in
the environments where the storage cost is more important than the
energy cost, the user gives a smaller value of $\alpha$. In this
paper, in order to control the trade-off between the two costs, we
define the reference value of $\alpha$, denoted as $\alpha_0$, which
makes the importance of the two costs equal. This $\alpha_0$ is the
value for balancing between the two costs which use different
scales, and is used as an example to determine the appropriate value
of $\alpha$ for a given application. Eq.(\ref{eq:std_alpha_eqn})
shows the definition of $\alpha_0$:

\begin{equation}
\alpha_0 = \frac{\mbox{\it the maximum possible total amount of
memory used}}{\mbox{\it the maximum possible total amount of data
sent}} \label{eq:std_alpha_eqn}
\end{equation}

\noindent In this equation, the denominator represents the total
amount of data sent from sensor nodes when every node stores
\emph{only one} query merged from all the original queries, and the
numerator represents the total amount of memory used for storing
queries into sensor nodes when every node stores all the original
queries. That is, $\alpha_0$ is the result of dividing the worst
case memory usage amount by the worst case data transmission amount.

In Eq.(\ref{eq:weightedSum}), the total memory usage amount is
determined by the number of queries stored in the nodes at each
tier, and the total data transmission amount is determined by the
amount of data sent at each tier based on the queries. We first
introduce the notion of the \emph{merge rate} in order to formulate
the number of queries stored in sensor nodes at each tier. We use it
as the optimization parameter for the Weighted\_Sum. The merge rate
is defined as the ratio of the memory usage amounts of two nodes at
adjacent tiers, as shown in Eq.(\ref{eq:mergeRate}).

\begin{eqnarray}
\mbox{\it merge\_rate}~&=&~\frac{\mbox{\it
the~number~of~queries~stored~at~a~node~at~the~$i^{th}$~tier}}{\mbox{\it
the~number~of~queries~stored~at~a~node~at~the~$(i-1)^{th}$~tier}}\nonumber\\%
&& \mbox{for all $2 \leq i \leq h$, where $h$ is the height of the
hierarchical sensor network, and}\nonumber\\%
&& \mbox{the server is at the first(highest) tier storing all the
original queries.}\label{eq:mergeRate}
\end{eqnarray}

According to the definition above, the merge rate has the value in
the range of 0 to 1. If the value is closer to 0, it means that more
queries are merged. On the other hand, if the value is closer to 1,
it means that fewer queries are merged. That is, the number of
queries stored in a node at each tier is determined by the merge
rate. For example, if the merge rate is 0, our approach is
equivalent to the centralized approach and if 1, it is equivalent to
the distributed approach.

Next, we introduce the notion of cover to formulate the amount of
data sent at each tier. The \emph{cover} is defined as the ratio of
the size of the domain space filled by all query regions over the
size of the entire domain space. In order to obtain the exact amount
of data transmission, we need additional information at each tier
such as the selectivity of each merged query and the size of each
dead region caused by query merging. This kind of information,
however, is affected significantly by the application environment
including the data and query distributions, making it difficult to
obtain exact information at the time of designing the network. Thus,
in this paper, we use an approximate model of the cover instead.
Definition \ref{def:cover} shows the definition of the cover of a
query set $Q$.

\begin{definition}[The cover of a query set Q]\label{def:cover}{\rm
For a \\given query set $Q=\{~q_1,~q_2,~\cdots,~q_n~\}$, its cover
\emph{cover($Q$)} is defined as:
\begin{eqnarray}
\mbox{\it cover}(Q)=\frac{\parallel
\Phi(q_1)\bigoplus\cdots\bigoplus\Phi(q_n)\parallel}{\parallel D
\parallel}\nonumber
\end{eqnarray}
\vspace{-0.5cm}
\begin{align}
&\mbox{for }q_i,q_j\in Q(1\leq i < j \leq n),\nonumber\\
&\mbox{where\,$D$\,is\,the\,domain\,space}, \nonumber\\
&\Phi(q_i)\mbox{ is the
region of the query }q_i,\nonumber\\
&\Phi(q_i)\bigoplus\Phi(q_j)\mbox{ represents the union of the two regions }\nonumber\\
&\Phi(q_i)\mbox{ and }\Phi(q_i),\mbox{ and }\nonumber\\
&\parallel\cdot\parallel\mbox{ denotes the size of the given
region.}
\end{align}
} \hfill\fbox{}%
\end{definition}

\begin{figure}[h!]
  \centering
  \includegraphics[width=2.5in]{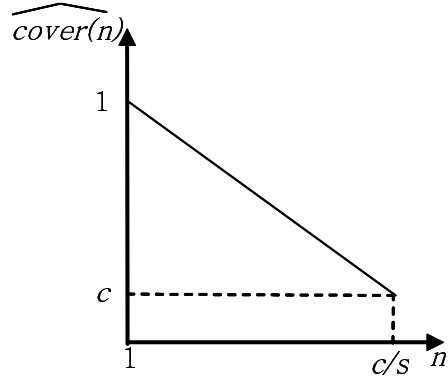}\\
  \caption{The cover model.}\label{fig:cover_func_eqn}
\end{figure}

Assuming that queries are uniformly distributed in the domain space,
$cover(Q)$ can be approximated at each tier as follows. Let $n$
denote the number of merged queries, $s$ denote the average
selectivity of the set of the original queries, and $c$ denote the
cover of the set of the original queries, then $\widehat{cover}(n)$
in Figure\,\ref{fig:cover_func_eqn} is an approximation of
$cover(Q)$.

$\widehat{cover}(n)$ has the following properties: (1) If $n$~=~1,
$\widehat{cover}(n)$ equals 1; (2) As $n$ increases,
$\widehat{cover}(n)$ decreases becoming $c$ when $n$=$\frac{c}{s}$.
That is,
$\widehat{cover}(n)\leq\widehat{cover}(n-1)\leq\cdots\leq$$\widehat{cover}(1)=1.$

These properties are from fact that the proposed merge method is
based on MBR. Since the region of a merged query is represented by
an MBR enclosing the regions of queries that are merged, the size of
the region of the merged query is always greater than or equal to
the size of the region resulting from the union of the query regions
that are merged. Thus, as the query merging proceeds, the number of
merged queries $n$ decreases, but the size of the region that is
equivalent to the union of merged queries increases. In this paper
we have assumed an environment in which we process a large number of
queries with the uniform distribution, and thus, we assume that the
cover of the merged query is 1. Even though this property does not
guarantee the linearity of $\widehat{cover}(n)$, in order to make
the model simple, we assume that the cover linearly increases as $n$
decreases, and then, estimate the theoretical number of queries for
which the cover is completely filled without overlap region as
\begin{math}\frac{\displaystyle \textsl{the cover of original
queries }}{\displaystyle \textsl{the average selectivity of original
queries}}\end{math}.

\subsection{Optimization}

\noindent In this subsection, we first formulate Weighted\_Sum using
the merge\_rate and the cover model explained in Section 4.1, and
then, analytically obtain the optimal merge rate -- the merge rate
that minimizes Weighted\_Sum. Table\,1 shows the notation used in
this section. For ease of exposition, we assume that each sensor
node generates only one data element per unit time.

\renewcommand{\baselinestretch}{1.35}
\begin{table}[h!]
\begin{center}
\label{table:parameter_symbol_table} \caption{The notation.}
\vspace*{0.2cm}
\begin{tabular}{|c|l|}\hline
        \hspace*{0.1cm}\textbf{Symbol}&
        \hspace*{0.1cm}\textbf{Definition} \\
    \hline\hline
    $N_Q$&The number of original queries \\
    \hline
    $c$&The cover of original queries \\
    \hline
    $s$&The average selectivity of original queries \\
    \hline
    $d$&The dimension of original queries \\
    \hline
    $h$&The height of a hierarchical sensor network \\
    \hline
    $f$&The fanout of a hierarchical sensor network \\
    \hline
    $Size_{de}$&The size of a data element \\
    \hline
    $m$&The merge rate \\
    \hline
\end{tabular}
\vspace*{-0.5cm}
\end{center}
\end{table}
\renewcommand{\baselinestretch}{2.0}

The total\_transmission(i.e., the total amount of data sent per unit
time) is formulated as follows (refer to Appendix A for details):

\begin{eqnarray}
total\_transmission&=&\sum_{i=2}^h(Size_{de}~\cdot~f^{i-1}~\cdot~\sum_{j=2}^i(-a~\cdot~m^{j-1}~\cdot~N_Q+b))\nonumber\\
\mbox{where}~a&=&\frac{\displaystyle s~\cdot~(1~-~c)}{\displaystyle
c~-~s},~b~=~1~+~a \label{eq:total_trans_eqn}
\end{eqnarray}

\newpage\noindent The total\_storage(i.e., the total amount of memory used)
is formulated as follows~(refer to Appendix A for details):
\begin{eqnarray}
total\_storage~=~\sum_{i=2}^h(~2~\cdot~Size_{de}~\cdot~f^{i-1}~\cdot~N_Q~\cdot~m^{i-1}) 
\label{eq:total_storage_eqn}
\end{eqnarray}

\noindent From Eq.(\ref{eq:total_trans_eqn}) and
Eq.(\ref{eq:total_storage_eqn}), Weighted\_Sum is formulated as
follows.
\begin{eqnarray}
Weighted\_Sum&=&\alpha~\cdot~total\_transimission~+~total\_storage\nonumber\\
             &=&\alpha~\cdot~Size_{de}~\cdot~\sum_{i=2}^h(f^{i-1}~\cdot~[~\sum_{j=2}^i(-a~\cdot~m^{j-1}~\cdot~N_Q+b))\nonumber\\
             &+&~2~\cdot~N_Q~\cdot~m^{i-1}~])\nonumber\\
\mbox{where}~ a&=&\frac{\displaystyle s~\cdot~(1~-~c)}{\displaystyle
c~-~s},~b~=~1~+~a \label{eq:ws_eqn}
\end{eqnarray}

In order to obtain the optimal merge rate, we take the derivative of
the Weighted\_Sum formula with respect to $m$ and compute the roots
from the derivative formula. Then, we substitute each root for $m$
in the Weighted\_Sum formula and find the root that minimizes the
computed Weighted\_Sum. We use Maple\cite{Map08}, a mathematics
software tool, for this computation.

%
%
\section{Performance
evaluation}\label{sec:PerformanceEvaluation} \vspace*{-0.30cm}

\subsection{Experimental data and environments}

\noindent We use two sets of experiments. In the first set, we show
the accuracy of the proposed cost model as the parameters are
varied. In the second set, we show the merit of our progressive
approach over the iterative approach proposed by Xiang et
al.\cite{Xia06} in terms of the total cost (i.e., Weighted\_Sum) of
query processing as the parameters are varied. A common set of seven
parameters are used in both sets of experiments: the scale factor
$\alpha$ for controlling the ``importance'' between the amount of
data transmission and the amount of memory usage, the cover of
original queries $c$, the average selectivity of original queries
$s$, the dimension of original queries $d$, the height of the sensor
network $h$, the fanout of the sensor network $f$, and merge rate
$m$. We use Weighted\_Sum as both the accuracy and the performance
measure.

We use the same data and query sets in both sets of experiments. We
randomly generate synthetic queries and data with the uniform
distribution. Here, ``uniform'' means that the locations of the
queries (or the data elements) are set randomly in the query space
(or the domain space). We generate queries with the same width in
all domains(i.e., hypercubes) in two alternative ways: either by
controlling the number of original queries or by controlling the
cover of original queries. The latter is used only in the
experiments for varying the cover of original queries, and the
former is used in all the other experiments. The reason we do not
control the number and the cover of the queries together is that
there is a dependency between the two values. That is, given a set
of random queries with a uniform distribution, if the number of
queries increases (with the query selectivity fixed) then the cover
also increases. This makes it impossible to generate a query set
with a uniform distribution when both number and cover are
controlled at the same time.

In the first set of experiments, we experimentally evaluate the
accuracy of our model for estimating the optimal merge rate that
minimizes the weighted sum of the storage cost and the energy cost
(i.e., Eq.(\ref{eq:weightedSum})). We first analytically compute the
estimated optimal merge rate as explained in Section\,4.2. Next, we
experimentally find the actual optimal merge rate. Finally, we
compare the two optimal merge rates. Table\,2 summarizes the
experiments and the parameters used.

\renewcommand{\baselinestretch}{1.2}
\begin{table}[h!]
\begin{center}
\label{table:exp_parameter_table2} \caption{Experiments and
parameters used for showing the accuracy of the cost model.}
\vspace*{0.2cm}
\begin{tabular}{|p{2cm}|p{4cm}|p{1cm}|p{6cm}|}\hline
    \multicolumn{2}{|p{6cm}|}{\begin{center}{\vspace*{-0.5cm}Experiments}\vspace*{-0.5cm}\end{center}}
    &
    \multicolumn{2}{|p{7cm}|}{\begin{center}{\vspace*{-0.5cm}Parameters}\vspace*{-0.5cm}\end{center}}\\
    \hline\hline
    Experiment\,1 &
    accuracy & $h$ & 4 \\ \cline {3-4}
    &as $\alpha$ is varied & $f$ & 8 \\ \cline {3-4}
    & & $\alpha$ & $\alpha_0\cdot10^{-2}$,$\alpha_0\cdot10^{-1}$,~$\alpha_0$,~$\alpha_0\cdot10^{1}$,~$\alpha_0\cdot10^{2}$ \\ \cline {3-4}
    & & $s$ & $10^{-4}$ \\ \cline {3-4}
    & & $d$ & 2 \\
    \hline
    Experiment\,2 &
    accuracy & $h$ & 4 \\ \cline {3-4}
    &as $c$ is varied  & $f$ & 8 \\ \cline {3-4}
    & & $\alpha$ & $\alpha_0$ \\ \cline {3-4}
    & & $d$ & 2 \\ \cline {3-4}
    & & $c$ & 0.01, 0.10, 0.99 \\
    \hline
    Experiment\,3 &
    accuracy & $h$ & 4 \\ \cline {3-4}
    &as $s$ is varied & $f$ & 8 \\ \cline {3-4}
    & & $\alpha$ & $\alpha_0$ \\ \cline {3-4}
    & & $s$ & $10^{-5}$,$10^{-4}$,$10^{-3}$ \\ \cline {3-4}
    & & $d$ & 2 \\
    \hline
    Experiment\,4 &
    accuracy & $h$ & 3, 4, 5 \\ \cline {3-4}
    &as $h$ is varied & $f$ & 8 \\ \cline {3-4}
    & & $\alpha$ & $\alpha_0$ \\ \cline {3-4}
    & & $s$ & $10^{-4}$ \\ \cline {3-4}
    & & $d$ & 2 \\
    \hline
    Experiment\,5 &
    accuracy & $h$ & 4 \\ \cline {3-4}
    &as $f$ is varied & $f$ & 2, 4, 8, 16 \\ \cline {3-4}
    & & $\alpha$ & $\alpha_0$ \\ \cline {3-4}
    & & $s$ & $10^{-4}$ \\ \cline {3-4}
    & & $d$ & 2 \\
    \hline
    Experiment\,6 &
    accuracy & $h$ & 4 \\ \cline {3-4}
    &as $d$ is varied & $f$ & 8 \\ \cline {3-4}
    & & $\alpha$ & $\alpha_0$ \\ \cline {3-4}
    & & $s$ & $10^{-4}$ \\ \cline {3-4}
    & & $d$ & 1, 2, 3 \\
    \hline
\end{tabular}
\vspace*{-0.5cm}
\end{center}
\end{table}
\renewcommand{\baselinestretch}{2.0}

In the second set of experiments, we compare the performance merit
of our progressive approach with the iterative approach proposed by
Xiang et al.\cite{Xia06}. We measure Weighted\_Sum while varying
parameters explained above. Here, in our approach, we use the
estimated optimal merge rate measuring Weighted\_Sum while varying
parameters explained above. Table\,3 summarizes the experiments and
the parameters used.

\renewcommand{\baselinestretch}{1.2}
\begin{table}[h!]
\begin{center}
\label{table:exp_parameter_table2} \caption{Experiments and
parameters used for showing the performance merit of our approach.}
\vspace*{0.2cm}
\begin{tabular}{|p{2.5cm}|p{3.5cm}|p{1cm}|p{6cm}|}\hline
    \multicolumn{2}{|p{6cm}|}{\begin{center}{\vspace*{-0.5cm}Experiments}\vspace*{-0.5cm}\end{center}}
    &
    \multicolumn{2}{|p{7cm}|}{\begin{center}{\vspace*{-0.5cm}Parameters}\vspace*{-0.5cm}\end{center}}\\
    \hline\hline
    Experiment\,7 &
    comparison of& $h$ & 4 \\ \cline {3-4}
    &the performance & $f$ & 8 \\ \cline {3-4}
    &as $\alpha$ is varied& $\alpha$ & $\alpha_0\cdot10^{-2}$,$\alpha_0\cdot10^{-1}$,~$\alpha_0$,~$\alpha_0\cdot10^{1}$,~$\alpha_0\cdot10^{2}$ \\ \cline {3-4}
    & & $s$ & $10^{-4}$ \\ \cline {3-4}
    & & $d$ & 2 \\
    \hline
    Experiment\,8 &
    comparison of & $h$ & 4 \\ \cline {3-4}
    &the performance & $f$ & 8 \\ \cline {3-4}
    &as $c$ is varied & $\alpha$ & $\alpha_0$ \\ \cline {3-4}
    & & $d$ & 2 \\ \cline {3-4}
    & & $c$ & 0.01, 0.10, 0.99 \\
    \hline
    Experiment\,9 &
    comparison of & $h$ & 4 \\ \cline {3-4}
    &the performance & $f$ & 8 \\ \cline {3-4}
    &as $s$ is varied & $\alpha$ & $\alpha_0$ \\ \cline {3-4}
    & & $s$ & $10^{-3}$,$10^{-4}$,$10^{-5}$ \\ \cline {3-4}
    & & $d$ & 2 \\
    \hline
    Experiment\,10 &
    comparison of & $h$ & 3, 4, 5 \\ \cline {3-4}
    &the performance & $f$ & 8 \\ \cline {3-4}
    &as $h$ is varied & $\alpha$ & $\alpha_0$ \\ \cline {3-4}
    & & $s$ & $10^{-4}$ \\ \cline {3-4}
    & & $d$ & 2 \\
    \hline
    Experiment\,11 &
    comparison of & $h$ & 4 \\ \cline {3-4}
    &the performance & $f$ & 2, 4, 8, 16 \\ \cline {3-4}
    &as $f$ is varied & $\alpha$ & $\alpha_0$ \\ \cline {3-4}
    & & $s$ & $10^{-4}$ \\ \cline {3-4}
    & & $d$ & 2 \\
    \hline
    Experiment\,12 &
    comparison of & $h$ & 4 \\ \cline {3-4}
    &the performance & $f$ & 8 \\ \cline {3-4}
    &as $d$ is varied & $\alpha$ & $\alpha_0$ \\ \cline {3-4}
    & & $s$ & $10^{-4}$ \\ \cline {3-4}
    & & $d$ & 1, 2, 3 \\
    \hline
\end{tabular}
\vspace*{-0.5cm}
\end{center}
\end{table}
\renewcommand{\baselinestretch}{2.0}

All experiments have been conducted using a Linux-Redhat system with
a 4 GHz processor and 1 Gbytes of main memory. Since it is difficult
to build an actual large-scale sensor network and change its
configuration as we need, we conduct the experiments using a
simulator program as commonly used in sensor networks-related
database research\cite{Li03,Mad05,Xia06}. We have implemented the
simulator program using C. Table\,4 summarizes the notation used in
the next section to discuss the experimental results.

\renewcommand{\baselinestretch}{1.35}
\begin{table}[h!]
\begin{center}
\label{table:exp_definition_table} \caption{Notation for explaining
experiments.} \vspace*{0.2cm}
\begin{tabular}{|p{2.5cm}|p{12.5cm}|}\hline
        \hspace*{0.1cm}\textbf{Symbol}&
        \hspace*{0.1cm}\textbf{Definition} \\
    \hline\hline
    $m_{opt\_act}$ & The actual optimal merge rate measured \\
    \hline
    $m_{opt\_est}$ & The estimated optimal merge rate obtained using the analytical model \\
    \hline
    $w_{opt\_act}$ & Weighted\_Sum measured using $m_{opt\_act}$ \\
    \hline
    $w_{opt\_est}$ & Weighted\_Sum measured using $m_{opt\_est}$ \\
    \hline
    $ratio_m$ & The ratio of $m_{opt\_act}$ to $m_{opt\_est}$~=~$\frac{\displaystyle m_{opt\_act}}{\displaystyle m_{opt\_est}}$ \\
    \hline
    $ratio_w$ & The ratio of $w_{opt\_act}$ to $w_{opt\_est}$~=~$\frac{\displaystyle w_{opt\_act}}{\displaystyle w_{opt\_est}}$ \\
    \hline
    $gain_w$ & $\frac{\displaystyle \textsl{Weighted\_Sum measured using Xiang et al.'s interative approach }}{\displaystyle w_{opt\_est}}$ \\
    \hline
\end{tabular}
\vspace*{-0.5cm}
\end{center}
\end{table}
\renewcommand{\baselinestretch}{2.0}

\subsection{Experimental results}

\subsubsection{Accuracy of the cost model}

\begin{figure}[h!]
  \centering
  \includegraphics[width=6.3in]{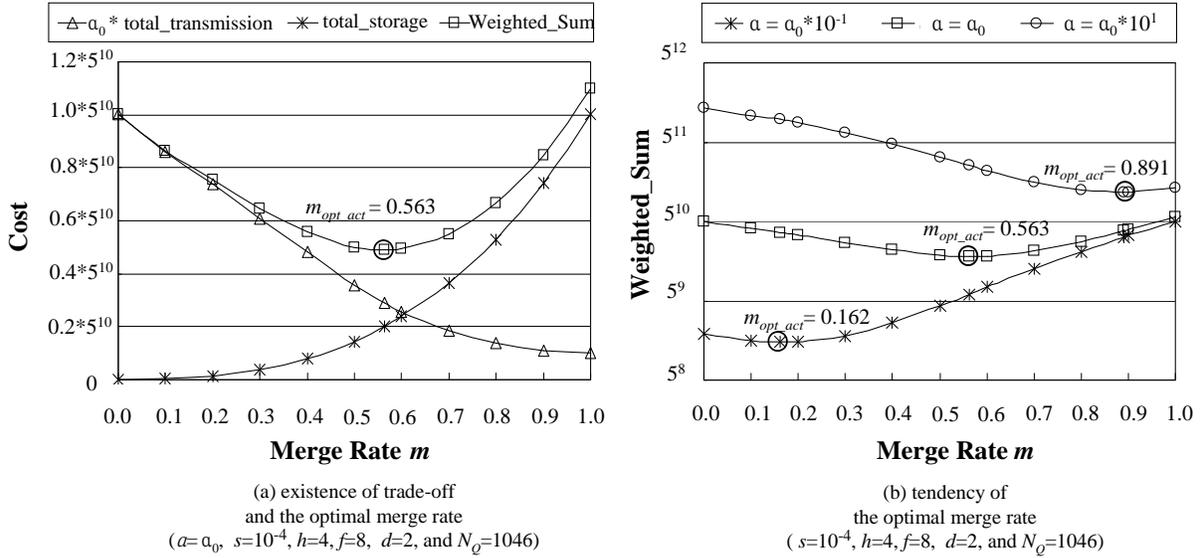}\\
  \caption{The existence of trade off and the optimal merge rate.}\label{fig:exp_opt1}
\end{figure}

\noindent \textbf{Experiment 0: existence of the trade-off
and the optimal merge rate}\\
\noindent Figure\,\ref{fig:exp_opt1}(a) shows the trade-off between
the total storage cost and the total transmission cost (i.e., energy
cost) as $m$ is varied. Here, we measure Weighted\_Sum for 1046
randomly generated queries(i.e., $N_Q$=1046). Hereafter, we use
$N_Q$=1046 unless we explicitly specify the value. As explained in
Section\,3.1, the transmission cost (i.e.,
$\alpha_0\cdot$total\_transmission) has a tendency to decrease as
$m$ increases. The storage cost has a tendency to increases as $m$
does. Thus, a value of $m$ that minimizes the weighted sum exists as
shown in Figure\,\ref{fig:exp_opt1}(a).
Figure\,\ref{fig:exp_opt1}(b) shows the trend of the actual optimal
merge rate as $m$ is varied. We observe that the optimal merge rate
has a tendency to increase as $\alpha$ does.


\begin{figure}[h!]
  \centering
  \includegraphics[width=3.7in]{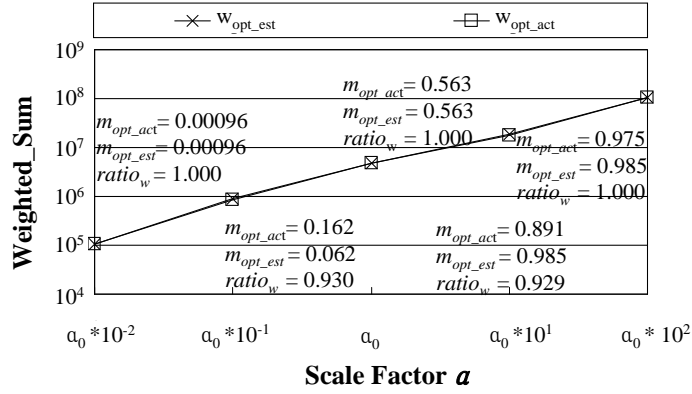}\\
  \caption{Optimal merge rate as $\alpha$ is varied($s$=$10^{-4}$, $h$=4, $f$=8, $d$=2, and $N_Q$=1046).}\label{fig:exp_alpha_merging_rate}
\end{figure}

\vspace{0.2cm}\noindent \textbf{Experiment 1: accuracy as $\alpha$ is varied}\\
Figure\,\ref{fig:exp_alpha_merging_rate} shows experimental results
as $\alpha$ is varied. We have different optimal merge rates for
different scale factors as shown in this figure. From
Figure\,\ref{fig:exp_alpha_merging_rate}, we see that $ratio_m$ is
0.905 to 2.619. Other than the value of 2.619 when $\alpha$ is
$\alpha_0\cdot10^{-1}$, $ratio_m$ is approximately 1.0 for all the
other values of $\alpha$. That is, the optimal merge\_rate measured
from the experimental data is almost the same as that obtained from
the analysis. Besides, we see that the value of $ratio_w$ is 0.929
to 1.0. That is, the values of Weighted\_Sum measured from the
experimental data are very close to those obtained from the
analysis. As we see from the result of this experiment, as $\alpha$
increases, the weight of the total transmission cost increases
relative to the weight of the total storage cost and, thus, the
optimal value is determined toward reducing the total transmission
cost -- toward making the optimal merge\_rate close to 1.

\begin{figure}[h!]
  \centering
  \includegraphics[width=3.7in]{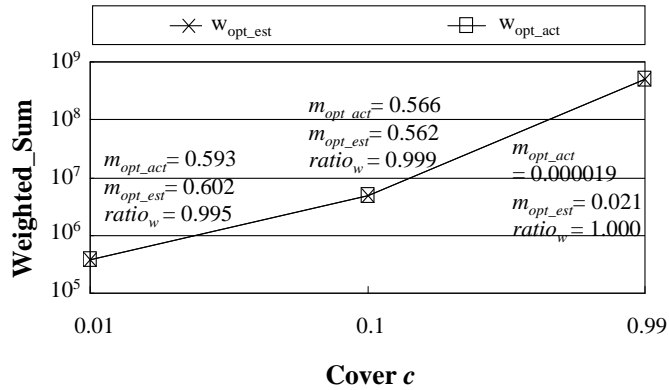}\\
  \caption{Optimal merge rate as $c$ is varied($\alpha=\alpha_0$, $s=10^{-4}$, $h$=4, $f$=8, and $d$=2).}\label{fig:exp_cover_merging_rate}
\end{figure}

\vspace{4pt}\noindent \textbf{Experiment 2: accuracy as $c$ is varied}\\
\noindent Figure\,\ref{fig:exp_cover_merging_rate} shows the
experimental results as the cover is varied. We use different query
sets for different covers (we use $N_Q$=101 when $c$=0.01,
$N_Q$=1046 when $c$=0.1, and $N_Q$=52685 when $c$=0.99). From
Figure\,\ref{fig:exp_cover_merging_rate}, we see that $ratio_m$ is
0.00092 to 1.008. Other than the value 0.00092 when the cover is
0.99, $ratio_w$ is approximately 1.0 for all the other values of the
cover. Besides, we see that the value of $ratio_w$ is 0.995 to 1.0.
That is, the Weighted\_Sum measured from the experiments is similar
to that obtained from the analysis. As the cover increases, the
difference between the maximum and the minimum amounts of data
transmission should decrease. Thus, reduction of total data
transmission cost have no significant influence on the total cost if
the cover increases. Hence, the optimal value is determined toward
reducing the total storage cost -- toward making the optimal merge
rate close to 0.

\begin{figure}[h!]
  \centering
  \includegraphics[width=3.7in]{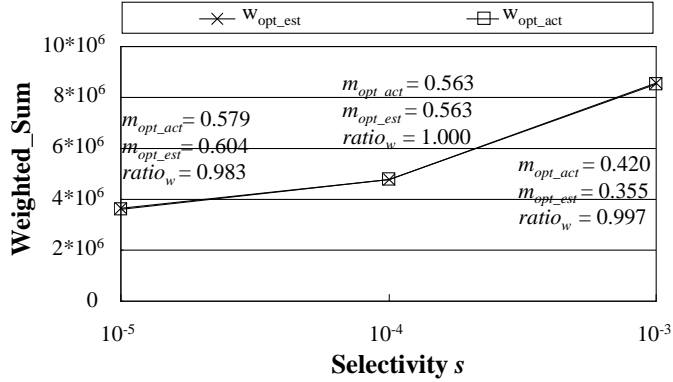}\\
  \caption{Optimal merge rate as $s$ is varied($\alpha=\alpha_0$, $h$=4, $f$=8, $d$=2, and $N_Q$=1046).}\label{fig:exp_selectivity_merging_rate}
\end{figure}

\vspace{4pt}\noindent\textbf{Experiment 3: accuracy as $s$ is varied}\\
\noindent Figure\,\ref{fig:exp_selectivity_merging_rate} shows the
experimental results as the selectivity is varied. From
Figure\,\ref{fig:exp_selectivity_merging_rate}, we see that
$ratio_m$ is 0.958 to 1.183 and $ratio_w$ is 0.983 to 1.0. The
increase of the selectivity is closely related to the increase of
the cover. That is, if the selectivity increases while the number of
queries is fixed, then the cover of the original queries increases
as well, and, thus, like the case of varying the cover, the optimal
value moves toward reducing the total storage cost -- toward making
the optimal merge rate close to 0.

\begin{figure}[h!]
  \centering
  \includegraphics[width=3.7in]{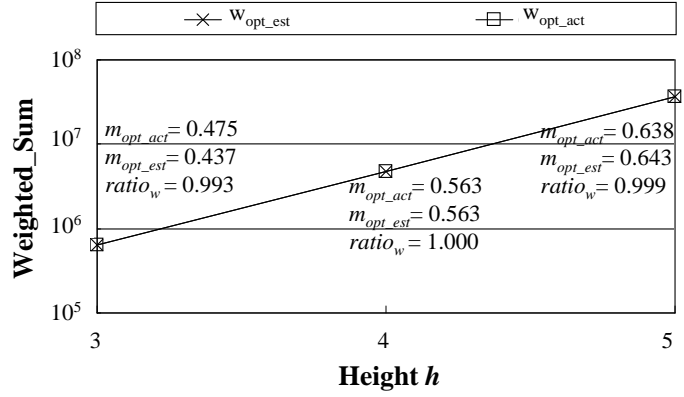}\\
  \caption{Optimal merge rate as $h$ is varied($\alpha=\alpha_0$, $s=10^{-4}$, $f$=8, $d$=2, and $N_Q$=1046).}\label{fig:exp_height_merging_rate}
\end{figure}

\vspace{4pt}
\noindent \textbf{Experiment 4: accuracy as $h$ is varied}\\
\noindent Figure\,\ref{fig:exp_height_merging_rate} shows the
experimental results as the height is varied. We see that $ratio_m$
is 0.993 to 1.088 and $ratio_w$ is 0.993 to 1.0. When the height of
the sensor network increases, the data transmission cost increases
faster than the memory usage cost. This stems from the fact that the
data sent are accumulated at each tier. Thus, the optimal value
moves toward reducing the total data transmission cost -- toward
making the optimal merge rate close to 1.

\begin{figure}[h!]
  \centering
  \includegraphics[width=3.7in]{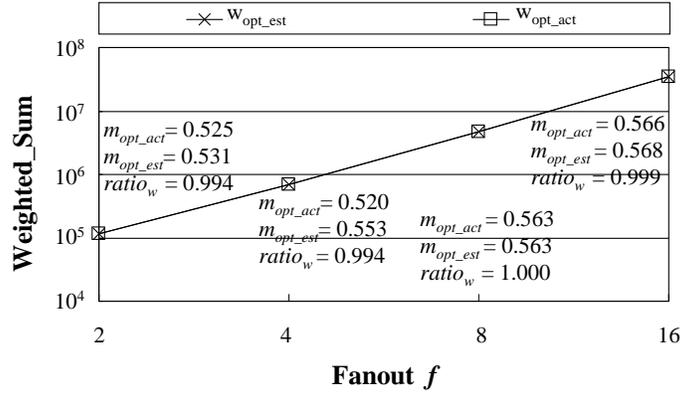}\\
  \caption{Optimal merge rate as $f$ is varied($\alpha=\alpha_0$, $s=10^{-4}$, $h$=4, $d$=2, and $N_Q$=1046).}\label{fig:exp_fanout_merging_rate}
\end{figure}

\vspace{4pt}
\noindent \textbf{Experiment 5: accuracy as $f$ is varied}\\
\noindent Figure\,\ref{fig:exp_fanout_merging_rate} shows the
experimental results as the fanout is varied. We see that $ratio_m$
is 0.993 to 1.088 and $ratio_w$ is 0.994 to 1.0. 

\begin{figure}[h!]
  \centering
  \includegraphics[width=3.7in]{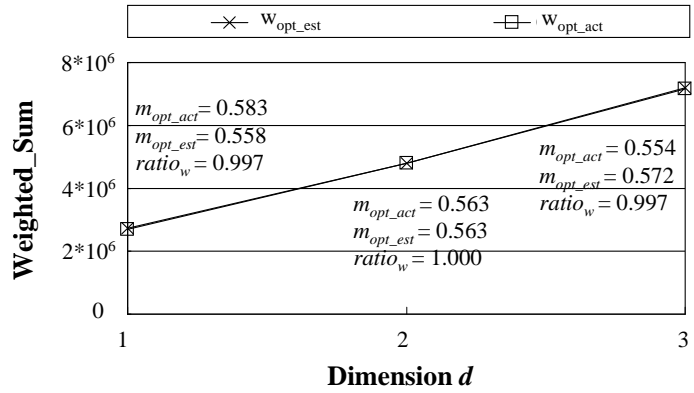}\\
  \caption{Optimal merge rate as $d$ is varied($\alpha=\alpha_0$, $s=10^{-4}$, $h$=4, $f$=8, and $N_Q$=1046).}\label{fig:exp_dim_merging_rate}
\end{figure}

\vspace{4pt}
\noindent \textbf{Experiment 6: accuracy as $d$ is varied}\\
\noindent Figure\,\ref{fig:exp_dim_merging_rate} shows the
experimental results as the dimension is varied. We see that
$ratio_m$ is 0.993 to 1.088 and $ratio_w$ is 0.997 to 1.0. 

\subsubsection{Performance merit of our approach}

\begin{figure}[h!]
  \centering
  \includegraphics[width=3.7in]{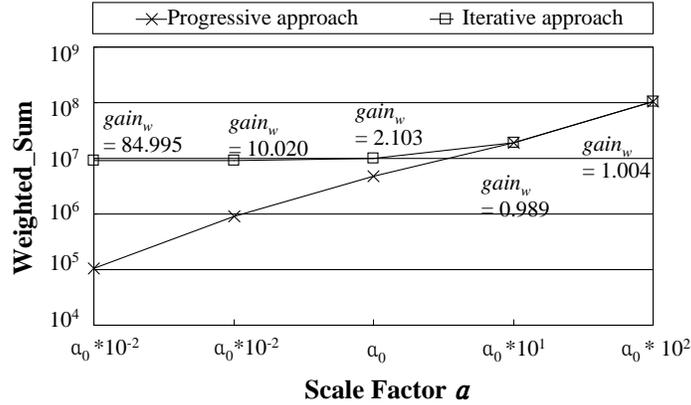}\\
  \caption{The performance of progressive approach and iterative approach as $\alpha$ is varied($s=10^{-4}$, $h$=4, $f$=8, $d$=2, and $N_Q$=1046).}\label{fig:exp_alpha_xiang}
\end{figure}

\noindent \textbf{Experiment 7: performance as $\alpha$ is
varied}\\
\noindent Figure\,\ref{fig:exp_alpha_xiang} shows the experimental
result as $\alpha$ is varied. Here, we have different merge rates
estimated for different scale factors (we use $m_{opt\_est}$=0.00096
when $\alpha$=$\alpha_0\cdot10^{-2}$, $m_{opt\_est}$=0.062 when
$\alpha$=$\alpha_0\cdot10^{-1}$, $m_{opt\_est}$=0.563 when
$\alpha$=$\alpha_0$, $m_{opt\_est}$=0.985 when
$\alpha$=$\alpha_0\cdot10^{1}$, and $m_{opt\_est}$=0.985 when
$\alpha$=$\alpha_0\cdot10^{2}$)). From this figure, we can see that
$gain_w$ is 0.989 to 84.995. Except for the value 0.989 when
$\alpha$ equals $\alpha_0~\cdot~10$, $gain_w$ is 1.004 to 84.995,
that is, Weighted\_Sum in the progressive approach is smaller than
Weighted\_Sum in the iterative approach. The exception happens due
to the fact that the cover model used in this paper (see
Figure\,\ref{fig:cover_func_eqn}) is an approximation of the cover
in the real environment, and this introduces some error between the
actual cost and the estimated cost. From these results, we see that
our approach greatly improves the performance over the approach
proposed by Xiang et al.\cite{Xia06} when memory usage is the
prevailing cost(i.e., $\alpha$ is small), while giving a competitive
performance when data transmission is the prevailing cost(i.e.,
$\alpha$ is large).

\vspace{-0.2cm}

\begin{figure}[h!]
  \centering
  \includegraphics[width=3.7in]{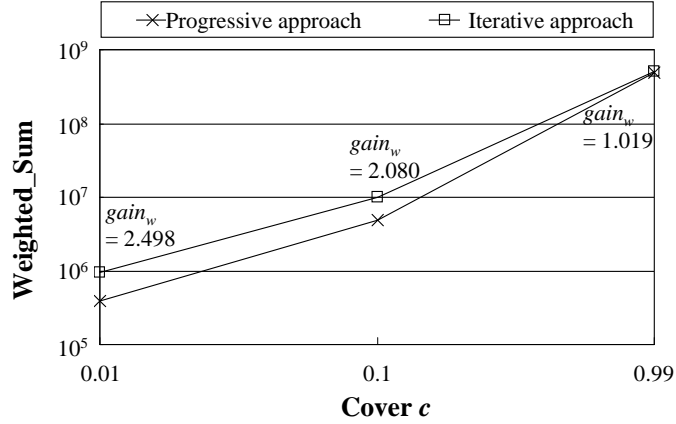}\\
  \caption{The performance of progressive approach and iterative approach as $c$ is varied($\alpha=\alpha_0$, $s=10^{-4}$, $h$=4, $f$=8, and $d$=2).}\label{fig:exp_cover_xiang}
\end{figure}

\vspace{0.2cm} \noindent \textbf{Experiment 8: performance as $c$ is
varied}\\
\noindent Figure\,\ref{fig:exp_cover_xiang} shows the experimental
result as the cover is varied. Here, we have different merge rates
estimated for different covers (we use $m_{opt\_est}$=0.602 when
$c$=0.01, $m_{opt\_est}$=0.562 when $c$=0.1, and
$m_{opt\_est}$=0.021 when $c$=0.99). We have different query sets
for different covers (we use $N_Q$=101 when $c$=0.01, $N_Q$=1046
when $c$=0.1, and $N_Q$=52685 when $c$=0.99). We see that $gain_w$
ranges from 1.019 to 2.498. This result shows that our approach
outperforms Xiang et al.'s approach in the entire range of the
cover. It also shows that, as the cover increases, the performance
benefit of our approach over Xiang et al.'s approach decreases. The
benefit of query merge with respect to the storage amount becomes
maximum when the cover approaches 1.0. In this case, all the
original queries are merged into one query in both our approach and
the Xiang et al.'s approach; as a result, the total transmission
amounts and the total storage amounts of the two approaches become
similar and, therefore, the weighted sums of the two approaches
become similar as well. Our proposed approach shows more performance
benefit when the cover of the original queries is smaller. The case
is more likely to happen in a real environment.

\begin{figure}[h!]
  \centering
  \includegraphics[width=3.7in]{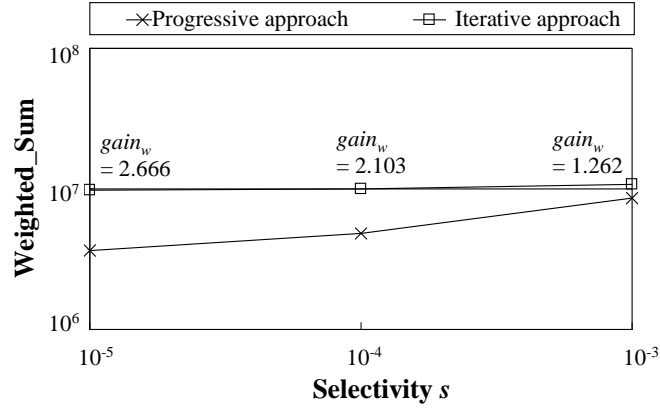}\\
  \caption{The performance of progressive approach and iterative approach as $s$ is varied($\alpha=\alpha_0$, $h$=4, $f$=8, $d$=2, and $N_Q$=1046).}\label{fig:exp_selectivity_xiang}
\end{figure}

\vspace{0.2cm} \noindent \textbf{Experiment 9: performance as $s$ is
varied}\\
\noindent Figure\,\ref{fig:exp_selectivity_xiang} shows the
experimental result as the average selectivity is varied. We have
different merge rates estimated for different selectivities (we use
$m_{opt\_est}$=0.604 when $s$=$10^{-5}$, $m_{opt\_est}$=0.563 when
$s$=$10^{-4}$, and $m_{opt\_est}$=0.355 when $s$=$10^{-3}$). We see
that $gain_w$ ranges from 1.262 to 2.666. This result shows that our
approach outperforms Xiang et al.'s approach in the entire range of
selectivity. It also shows that, as the selectivity increases, the
performance benefit of our approach decreases. As already mentioned
in the experiment that compares the optimal merge rates obtained
from the experimental data with those obtained from the analysis, if
the selectivity increases, then the cover increases as well causing
the decrease of performance benefit as we see in
Figure\,\ref{fig:exp_selectivity_xiang}. Thus, our proposed approach
shows more performance benefit when the selectivity of the original
queries is smaller.

\vspace{-0.2cm}

\begin{figure}[h!]
  \centering
  \includegraphics[width=3.7in]{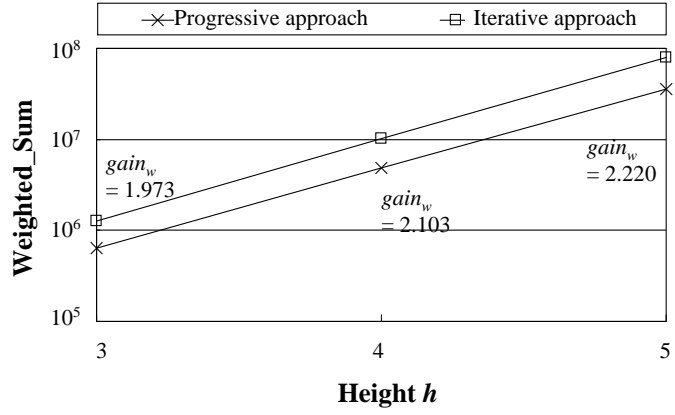}\\
  \caption{The performance of progressive approach and iterative approach as $h$ is varied($\alpha=\alpha_0$, $s=10^{-4}$, $f$=8, $d$=2, and $N_Q$=1046).}\label{fig:exp_height_xiang}
\end{figure}

\vspace{0.2cm} \noindent \textbf{Experiment 10: performance as $h$
is
varied}\\
\noindent Figure\,\ref{fig:exp_height_xiang} shows the experimental
result as height of the hierarchical sensor network is varied. We
have different merge rates estimated for different heights (we use
$m_{opt\_est}$=0.437 when $h$=3, $m_{opt\_est}$=0.563 when $h$=4,
and $m_{opt\_est}$=0.643 when $h$=5). We see that $gain_w$ ranges
from 1.973 to 2.220. This result shows that our approach outperforms
Xiang et al.'s approach in the entire range of the height. It also
shows that as the height increases, the performance benefit of our
approach increases slightly. The reason for this increase is that
the total storage amount in the iterative approach increases faster
than in the progressive approach as the height increases. That is,
in the iterative approach the same set of merged queries are stored
in all sensor nodes regardless of the tier whereas, in our
progressive approach, a smaller number of queries are stored as the
tier goes lower. Thus, our approach shows more performance benefit
when the height of the sensor network is larger.

\begin{figure}[h!]
  \centering
  \includegraphics[width=3.7in]{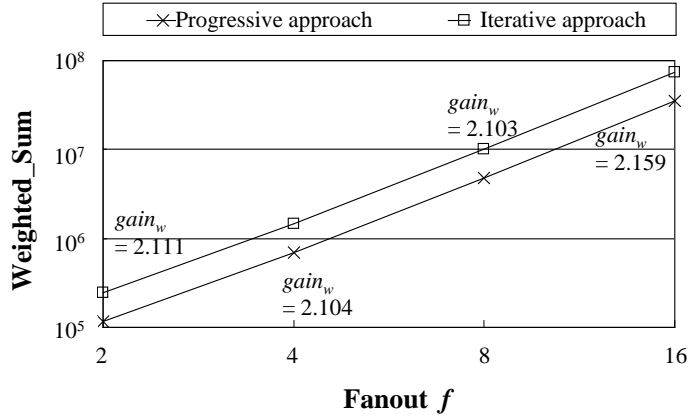}\\
  \caption{The performance of progressive approach and iterative approach as $f$ is varied($\alpha=\alpha_0$, $s=10^{-4}$, $h$=4, $d$=2, and $N_Q$=1046).}\label{fig:exp_fanout_xiang}
\end{figure}

\vspace{0.2cm} \noindent \textbf{Experiment 11: performance as $f$
is
varied}\\
\noindent Figure\,\ref{fig:exp_fanout_xiang} shows the experimental
result as the fanout of the sensor network is varied. We have
different merge rates estimated for different fanouts (we use
$m_{opt\_est}$=0.531 when $f$=2, $m_{opt\_est}$=0.553 when $f$=4,
$m_{opt\_est}$=0.563 when $f$=8, and $m_{opt\_est}$=0.568 when
$f$=16). In the result, $gain_w$ ranges from 2.103 to 2.159. We
observe that for all ranges of $f$, the performance of our approach
is better than that of the iterative approach.

\vspace{-0.5cm}

\begin{figure}[h!]
  \centering
  \includegraphics[width=3.7in]{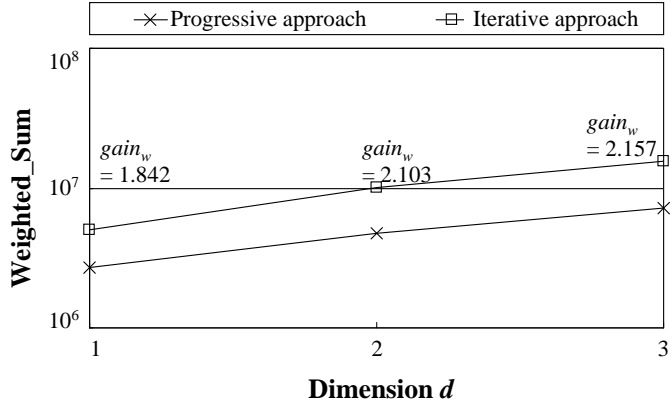}\\
  \caption{The performance of progressive approach and iterative approach as $d$ is varied($\alpha=\alpha_0$, $s=10^{-4}$, $h$=4, $f$=8, and $N_Q$=1046).}\label{fig:exp_dim_xiang}
\end{figure}

\noindent \textbf{Experiment 12: performance as $d$ is
varied}\\
\noindent Figure\,\ref{fig:exp_dim_xiang} shows the experimental
result as the dimension of a query is varied. We have different
merge rates estimated for different dimensions (we use
$m_{opt\_est}$=0.558 when $d$=1, $m_{opt\_est}$=0.563 when $d$=2,
and $m_{opt\_est}$=0.572 when $d$=3). In the result, $gain_w$ ranges
from 1.842 to 2.157. We observe that for all ranges of $f$, the
performance of our approach is better than that of the iterative
approach.

\vspace{0.2cm} In summary, the experimental results show that our
approach outperforms Xiang et al.'s approach by up to 84.995 times
as $\alpha$ is varied except when $\alpha$ is equal to
$\alpha_0~\cdot~10$. The results also show that our approach
outperforms Xiang et al.'s approach by up to 2.666 times as the
following other parameters are varied: the cover, average
selectivity, dimension of original queries, and the height, fanout
of the hierarchical sensor network.

%
%
\section{Conclusions}
\vspace*{-0.30cm}

\noindent In this paper, we have proposed \emph{progressive
processing} as a new approach to processing continuous range queries
in hierarchical sensor networks. The contribution of this paper are
summarized as follows.

First, we have proposed a progressive processing model that
considers the trade-off between energy and storage. This model takes
advantage of the characteristics of the hierarchical sensor networks
in which higher capability sensor nodes are deployed at a tier
closer to the server. It also has the advantage of reducing the cost
of building the network by reducing the storage cost at lower tier
nodes, which are larger in number. We also have presented query
merging and query processing algorithms for this model.

Second, based on the proposed model, we have proposed a method for
optimizing the total cost (formulated as the weighted sum of the
energy and storage costs) according to the given weight, and have
proposed a method for systematically building a hierarchical sensor
network that minimizes the total cost.

Third, we have verified the merit of the proposed approach through
extensive experiments. In the experiments for evaluating the
accuracy of the proposed cost model, the results show that the ratio
of the optimal cost measured over that obtained from the analytical
cost model is 0.929 to 1.0. From these results we see that a
hierarchical sensor network with near-optimal total cost can be
built using the proposed model. In the experiments for evaluating
the query processing performance, the results show that our approach
outperforms the approach proposed by Xiang et al.\cite{Xia06} by up
to 84.995 times. Moreover, if the height of the sensor network
increases, our approach shows a better performance than Xiang et
al.'s approach. Thus, we can see that our approach is suitable for a
\emph{large-scale} sensor network.

In conclusion, our approach provides a new framework for building a
large-scale hierarchical sensor network that efficiently processes a
large number of queries while considering the trade-off between the
energy consumed and the storage required.

For further work, we plan to improve the query processing model and
algorithms to consider different data and query distributions as
well as different query types such as aggregate queries.

%
%
\section{Acknowledgement}
\vspace*{-0.3cm}

This work was supported by the Korea Science and Engineering
Foundation (KOSEF) and the Korean Goverment (MEST) through the NRL
Program (No. R0A-2007-000-20101-0).

%
%

\section*{Appendix-A Derivation of Formula for
Total\_Transmission and Total\_Storage}

\textbf{\noindent Derivation of total\_transmission}

\noindent The total amount of data sent, denoted as
\emph{total\_transmission}, is the sum of the amounts of data sent
by all nodes at each tier while they are relayed to the server.
Eq.(\ref{eq:trans_sub1_eqn}) shows the formula for computing
total\_transmission.

\vspace*{-0.6cm}
\begin{eqnarray}
total\_transmission&=&\sum_{i=2}^h(Amt\_data_{i}~\cdot~\sum_{j=2}^i(c_j))\nonumber\\
\mbox{where}~c_j&=&\mbox{the~cover~of~merged~queries~stored~at~the~$j^{th}$~tier, and } \nonumber\\
Amt\_data_i&=&\mbox{the~amount~of~data~generated~by~the~sensor~nodes~at~the~$i^{th}$~tier}
\label{eq:trans_sub1_eqn}
\end{eqnarray}
\vspace*{-0.5cm}

In Eq.(\ref{eq:trans_sub1_eqn}), $c_j$ is formulated as follows
using the definition of the cover model (see
Figure\,\ref{fig:cover_func_eqn}) and the merge\_rate.
\begin{eqnarray}
c_j~=~\widehat{cover}(N_Q~\cdot~m^{j-1})\label{eq:trans_sub2_eqn}
\end{eqnarray}
where $N_Q~\cdot~m^{i-1}$ is the number of queries stored at the
$j^{th}$ tier (note $N_Q$ is the number of queries stored in the
server (at the $1^{st}$ tier) and $m$ is the merge rate between two
nodes in adjacent tiers (see
Table\,\ref{table:parameter_symbol_table})).
In the same Eq.(\ref{eq:trans_sub1_eqn}), $Amt\_data_i$ is
formulated as follows, based on the assumption that each sensor node
generates only one data element per unit time.

\begin{eqnarray}
Amt\_data_i&=&(\mbox{the~number~of~sensor~nodes~at~the~$i^{th}$~tier})~\cdot~(\mbox{the~size~of~a data~element})\nonumber\\
   &=&f^{i-1}~\cdot~Size_{de}   \label{eq:trans_sub3_eqn}
\end{eqnarray}
\noindent By substituting $c_j$ and $Amt\_data_i$ in
Eq.(\ref{eq:trans_sub1_eqn}) with those from
Eq.(\ref{eq:trans_sub2_eqn}) and Eq.(\ref{eq:trans_sub3_eqn}), we
can rewrite the formula for total\_transmission as follows.

\vspace*{-0.5cm}
\begin{eqnarray}
total\_transmission&=&\sum_{i=2}^h(Amt\_data_{i}~\cdot~\sum_{j=2}^i(c_j))\nonumber\\
                   &=&\sum_{i=2}^h(Size_{de}~\cdot~f^{i-1}~\cdot~\sum_{j=2}^i(c_j))\nonumber\\
                   &=&\sum_{i=2}^h(Size_{de}~\cdot~f^{i-1}~\cdot~\sum_{j=2}^i(\widehat{cover}(N_Q~\cdot~m^{j-1})))\nonumber\\
                   &=&\sum_{i=2}^h(Size_{de}~\cdot~f^{i-1}~\cdot~\sum_{j=2}^i(-a~\cdot~m^{j-1}~\cdot~N_Q+b))\nonumber\\
\mbox{where}~a&=&\frac{\displaystyle s~\cdot~(1~-~c)}{\displaystyle
c~-~s}\,~\mbox{and}\,~b~=~1~+~a\label{eq:trans_sub4_eqn}
\end{eqnarray}

\noindent \textbf{Derivation of total\_storage}

\noindent The total amount of memory used, denoted as
\emph{total\_storage}, is the sum of the amounts of memory used by
all nodes at all tiers. Eq.(\ref{eq:mem_sub1_eqn}) shows the formula
for computing total\_storage.

\begin{eqnarray}
total\_storage&=&\sum_{i=2}^h(Amt\_mem_i)\nonumber\\
\mbox{where}~Amt\_mem_i&=&\mbox{the~amount~of~memory needed to
store}\nonumber\\ &&\mbox{the merged
queries~in~all~sensor~nodes~at~the~$i^{th}$~tier}\label{eq:mem_sub1_eqn}
\end{eqnarray}

The number of merged queries stored in a sensor node at the $i^{th}$
tier is formulated as $N_Q~\cdot~m^{i-1}$ using
Eq.(\ref{eq:mem_sub1_eqn}) and Eq.(\ref{eq:trans_sub2_eqn}). Thus,
$Amt\_mem_i$ is formulated as in Eq.(\ref{eq:mem_sub2_eqn}).

\begin{eqnarray}
Amt\_mem_i&=&(\mbox{the~number~of~merged~queries~stored~in~all~sensor~nodes~at~the
~$i^{th}$~tier})\nonumber\\
   &\cdot&(\mbox{the amount~of~memory~needed~for~storing~one~query})\nonumber\\
   &=&f^{i-1}~\cdot~N_Q~\cdot~m^{i-1}~\cdot~2\cdot Size_{de} \label{eq:mem_sub2_eqn}
\end{eqnarray}
\newpage
\noindent By substituting $Amt\_mem_i$ from
Eq.(\ref{eq:mem_sub2_eqn}) into Eq.(\ref{eq:mem_sub1_eqn}), the
formula for total\_storage can be rewritten as follows.
\begin{eqnarray}
total\_storage&=&\sum_{i=2}^h(Amt\_mem_i)\nonumber\\
              &=&\sum_{i=2}^h(f^{i-1}~\cdot~N_Q~\cdot~m^{i-1}~\cdot~2~\cdot~Size_{de})\nonumber\\
              &=&\sum_{i=2}^h(~2~\cdot~Size_{de}~\cdot~f^{i-1}~\cdot~N_Q~\cdot~m^{i-1})
\label{eq:mem_sub3_eqn}
\end{eqnarray}
\vspace{0cm}
\end{document}